\documentclass[%
 aip,
 reprint,
 amsmath,amssymb,
]{revtex4-2}

\usepackage{graphicx}% Include figure files
\graphicspath{{Figures/}}
\usepackage{siunitx}
\usepackage{chemformula}
\usepackage{hyperref}% add hypertext capabilities
\begin{document}

\preprint{AIP/123-QED}

\title{Field emission mitigation studies in LCLS-II cavities via \textit{in situ} plasma processing}% Force line breaks with \\

\author{Bianca Giaccone}
 \email{giaccone@fnal.gov}
 \altaffiliation{Physics Department, Illinois Institute of Technology, 10 W 35th St, Chicago, IL 60616 USA}
\author{Martina Martinello}%
\author{Paolo Berrutti}
\author{Oleksandr Melnychuk}
\author{Dmitri A. Sergatskov}
\author{Anna Grassellino}
\affiliation{%
Fermi National Accelerator Laboratory, Kirk Rd and \\
Pine St, Batavia, IL 60510 USA
}%
\author{Dan Gonnella}
\author{Marc Ross}
\affiliation{%
SLAC National Accelerator Laboratory, 2575 Sand Hill Rd, Menlo Park, CA 94025 USA
}%

\author{Marc Doleans}
\affiliation{%
Oak Ridge National Laboratory, 1 Bethel Valley Rd, Oak Ridge, TN 37830 USA
}%

\author{John F. Zasadzinski}
\affiliation{Physics Department, Illinois Institute of Technology, 10 W 35th St, Chicago, IL 60616 USA}

\date{\today}% It is always \today, today,

\begin{abstract}
\noindent
Field emission is one of the factors that can limit the performance of superconducting radio frequency cavities. In order to reduce possible field emission in LCLS-II (Linac Coherent Light Source II), we are developing plasma processing for \SI{1.3}{\giga\hertz} 9-cell cavities. Plasma processing can be applied \textit{in situ} in the cryomodule to mitigate field emission related to hydrocarbon contamination present on the cavity surface. \\
In this paper, plasma cleaning was applied to single cell and 9-cell cavities, both clean and contaminated; the cavities were cold tested before and after plasma processing in order to compare their performance. It was proved that plasma cleaning does not negatively affect the nitrogen doping surface treatment; on the contrary, it preserves the high quality factor and quench field. Plasma processing was also applied to cavities with natural field emission or artificially contaminated. It was found that this technique successfully removes carbon-based contamination from the cavity iris and that it is able to remove field emission in a naturally field emitting cavity. Vacuum failure experiments were simulated on four cavities, and in some cases plasma processing was able to achieve an increase in performance.
\end{abstract}

\maketitle

%\tableofcontents

\section{\label{sec:intro}Introduction}
A collaboration among Fermi National Accelerator Laboratory (FNAL), SLAC National Accelerator Laboratory and Oak Ridge National Laboratory (ORNL) is working to develop plasma processing for LCLS-II \cite{stohr2011linac, Galayda_IPAC2014-TUOCA01} \SI{1.3}{\giga\hertz} nitrogen doped \cite{grassellino2013nitrogen, crawford2014joint, grassellino2015n, gonnella2018industrialization} cavities. LCLS-II (Linac Coherent Light Source II) is the LCLS XFEL (X-ray Free Electron Laser \cite{mcneil2010x, kao2020challenges}) upgrade, and will utilize a superconducting linear accelerator, along with other cutting-edge components, to produce an X-ray laser beam 1E4 times brighter than LCLS \cite{cornacchia1998linac, schoenlein2017linac}.

The scope of plasma processing is to be applied \textit{in situ} in LCLS-II cryomodules and help mitigate hydrocarbon-related field emission in 9-cell cavities.

Field emission (FE) is a phenomenon that limits the accelerating gradient at which a cavity can operate \cite{padamsee1998rf}; it consists of electron emission from regions of the cavity surface with intense applied electric field \cite{fowler1928electron}. The emitted electrons are accelerated by the electric field and impact on the cavity walls depositing heat and creating bremsstrahlung X-rays. These electrons can also interact with and disrupt the beam passing through the cavity. The X-rays produced by the field emission can cause radiation damage in the cryomodule's components, decreasing their lifetime.
Once FE is activated, it limits the cavity's accelerating field and it causes a degradation in quality factor, due to the additional dissipation introduced by the emitted electrons. If field emission is severe it can cause the cavity's thermal breakdown and it can also activate the beamline, causing induced radioactivity in the cavity.

Sources of field emission are contaminants (dust or metal particles) or cavity surface defects that cause a local enhancement of the FE current. In addition, the presence of few monolayers of hydrocarbons, or other adsorbate gases, on the cavity surface can further decrease the \ch{Nb} work function \cite{bagus2008work}, facilitating field emission. 
The origin of hydrocarbon contamination on the cavity inner surface is not completely understood; however, its presence has been reported in literature in multiple cases \cite{doleans2016situ, cao2013, pudasaini2020analysis} and carbon has been observed both in the form of adventitious and as a local contamination on the \ch{Nb} surface. Doleans et al report in \cite{doleans2016situ} that evidence of volatile hydrocarbon has been found through residual gas analysis on thermally cycled SNS (Spallation Neutron Source \cite{mason2006spallation}) cryomodules; they explain that these signals must originate from the released gases that were previously condensed on the cavity walls at cryogenic temperature or from species produced during accelerator operation by the interaction of electrons with the cavity surface contaminants.

Plasma processing can be used on superconducting radio frequency (SRF) cavities \textit{in situ} in the cryomodules to remove the hydrocarbon contamination and restore the niobium work function obtaining a decrease in field emission and a corresponding increase in the accelerating gradient. This technique was first applied to SRF cavities at ORNL, where Doleans et al developed plasma cleaning for SNS high beta \SI{805}{\mega\hertz} cavities \cite{doleans2016ignition}. Plasma processing has been applied to multiple SNS cryomodules, both offline and online, showing improvement in the accelerating gradient \cite{doleans2016plasma}.
Starting from SNS experience, a new method of plasma ignition for LCLS-II \SI{1.3}{\giga\hertz} TESLA-shaped \cite{aune2000superconducting} cavities has been developed by Berrutti et al at FNAL \cite{berrutti2018, berrutti2019plasma, giaccone_srf2019-frcab7}.
Studies on plasma ignition are being conducted also at the Institute of Modern Physics, CAS (Chinese Academy of Sciences) on half wave resonators (HWR). Wu et al used an experimental setup that replicates the cavity assembled in the cryomodule to study plasma ignition with \ch{Ar/O_2} gas mixture in HWR cavities \cite{wu2018situ} and studied the effect of plasma processing on a HWR cavity contaminated with methane gas \cite{wu2019cryostat}. Huang et al \cite{huang2019effect} applied helium and plasma processing to low beta HWR.

In this paper we present the results of plasma processing applied to multiple \SI{1.3}{\giga\hertz} cavities. The cavities were cold tested before and after plasma processing in order to compare their performance in terms of quality factor ($\mathrm{Q_0}$) and radiation versus accelerating field ($\mathrm{E_{acc}}$). A N-doped \cite{grassellino2013nitrogen} single cell was used for the first plasma processing test in order to study the possible effect of the plasma on the surface treatment. Afterwards, plasma cleaning was applied on two 9-cell cavities cavities with natural field emission (meaning FE of unknown source, not caused by a contamination intentionally introduced in the cavity). Given the results of these tests, it was decided to investigate the efficacy of plasma processing on cavities artificially contaminated with carbon-related sources or through vacuum failure simulations.

\section{\label{sec:experimentalsetup} Experimental System and Plasma parameters}
Plasma cleaning uses a glow discharge \cite{brown1966introduction, fitzpatrick2014plasma} ignited inside the cavity volume to remove the hydrocarbons from the niobium surface, restoring the \ch{Nb} work function and causing a decrease in field emission \cite{tyagi2014plasma}. An inert gas (neon) is injected into the cavity to ignite and sustain the plasma and a small percentage of \ch{O_2} is added to the \ch{Ne}. The oxygen molecules are dissociated in the plasma and the reactive oxygen binds with the hydrocarbons on the surface, creating volatile by-products that are easily pumped out of the cavity.

The glow discharge is ignited inside RF volume, cell by cell, using the cavity's resonant modes. Scope of plasma processing is to be used \textit{in situ} in the cryomodule, so this technique relies only on the hardware present in the cryomodule cavity's assembly. The SNS method for plasma ignition is the dual tone excitation \cite{doleans2016situ} and uses the fundamental pass-band. For LCLS-II \SI{1.3}{\giga\hertz} cavities, it is not possible to ignite the glow discharge using the fundamental power coupler (FPC). As Berrutti et al explain in \cite{berrutti2018}, the cavity quality factor $\mathrm{Q_0}$ and the FPC $\mathrm{Q_{ext}}$ are highly mismatched at room temperature \cite{lclsIIdesign}. Therefore, the approach taken here for LCLS-II cavities, is a new plasma ignition method, developed by Berrutti et al, using the higher order modes (HOMs) and the HOM couplers \cite{berrutti2019plasma}. Modes belonging to two dipole pass-bands are used to ignite the glow discharge in the central cell and to plasma process the entire cavity. The procedure is composed of two identical rounds; during each round all the cavity cells are plasma processed. Using the newly developed HOM ignition method, the glow discharge is ignited in the central cell and then immediately transferred to cell number 9, passing through adjacent cells. Once arrived in the desired cell, the RF driving frequency is tuned in order to increase the resonant peak's frequency shift and to maximize the plasma density \cite{berrutti2019plasma, doleans2013plasma, slater1946microwave}. The cell is processed for \SI{50}{\minute}, then the plasma density is decreased and the glow discharge is transferred to the adjacent cell using a combination of resonant dipole modes. The procedure is repeated until all the cavity cells, from $\#$ 9 to $\#$ 1, have been plasma processed. Once the first round is completed, the second identical round is performed.

Each cell is processed for a total of \SI{100}{\minute}. The duration of each round has been decided using the data collected by the Residual Gas Analyzer (RGA) assembled on the pumping system: it has been observed that peaks in the carbon-related signals are often present when the plasma is ignited in (or transferred to) a new cell. These peaks usually decrease to the background level in approximately \SI{30}{\minute} or less. During the second round of plasma processing, there is usually no increase in the \ch{C}-related signals.
The RGA is used during the entire plasma cleaning procedure to monitor the concentration of oxygen in the mixture and the byproducts of the reaction between \ch{O2} and the hydrocarbons on the cavity surface.

\begin{figure*}
	\centering
	{\includegraphics[width=0.95\textwidth]{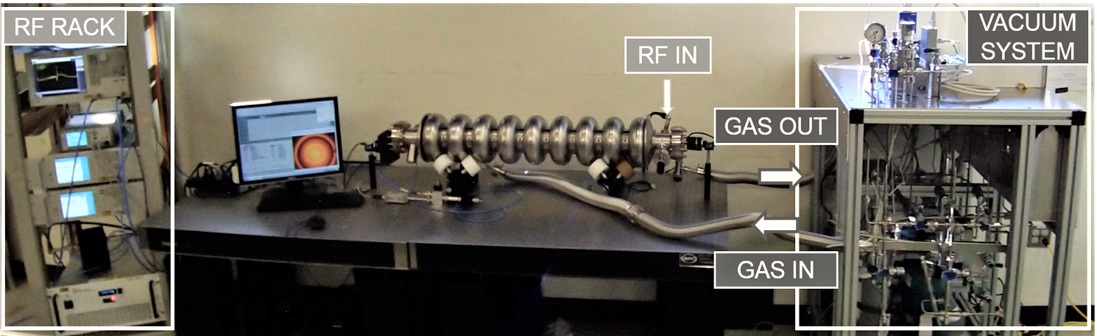}} \quad
	\caption[Experimental setup used for plasma processing]{Experimental setup used for plasma processing: RF rack on the left side, gas and vacuum system on the right, and cavity in the center. A portable cleanroom, not shown in the picture, has been added to the setup and placed around the cavity. All the connections between the cavity and the gas/vacuum system take place inside the cleanroom.}
	\label{fig:experimentalsetup}
\end{figure*}

Plasma processing on LCLS-II cavities is performed at room temperature, using a mixture of neon with approximately 1-1.5$\%$ oxygen, for a total gas pressure around $75\,$mTorr; the experimental setup in use at FNAL is shown in Fig. \ref{fig:experimentalsetup}.
The set of parameters currently used has been developed during the first plasma processing experiment on a 9-cell cavity, however it has not been optimized yet. Studies to identify the set of parameters (pressure, duration, oxygen percentage, plasma density/frequency shift) that maximizes the plasma efficiency are currently ongoing.

\subsection{\label{sec:permanentmarker} Removal study of \ch{C}-based contamination from cavity iris}
\begin{figure}
    \centering
    \includegraphics[width=1\columnwidth]{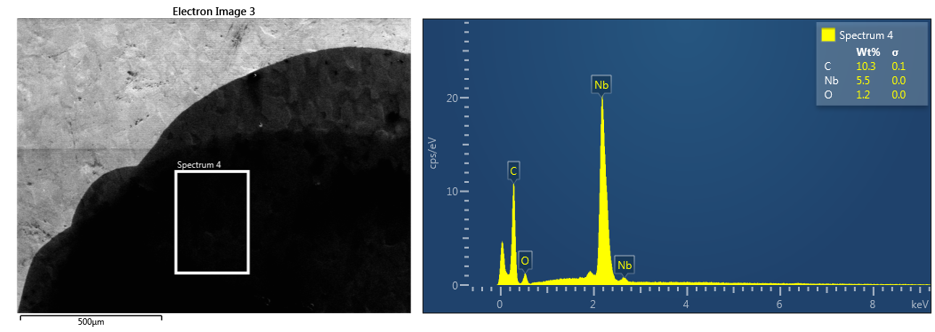}
    \caption[SEM/EDS analysis of permanent marker ink]{Permanent marker ink analyzed with Scanning Electron Microscopy (SEM) on the left and Energy Dispersive X-ray spectroscopy (EDS) on the right. A \ch{Nb} sample is used as substrate.}
    \label{fig:ink_EDS}
\end{figure}

\begin{figure}
    \centering
    \includegraphics[width=1\columnwidth]{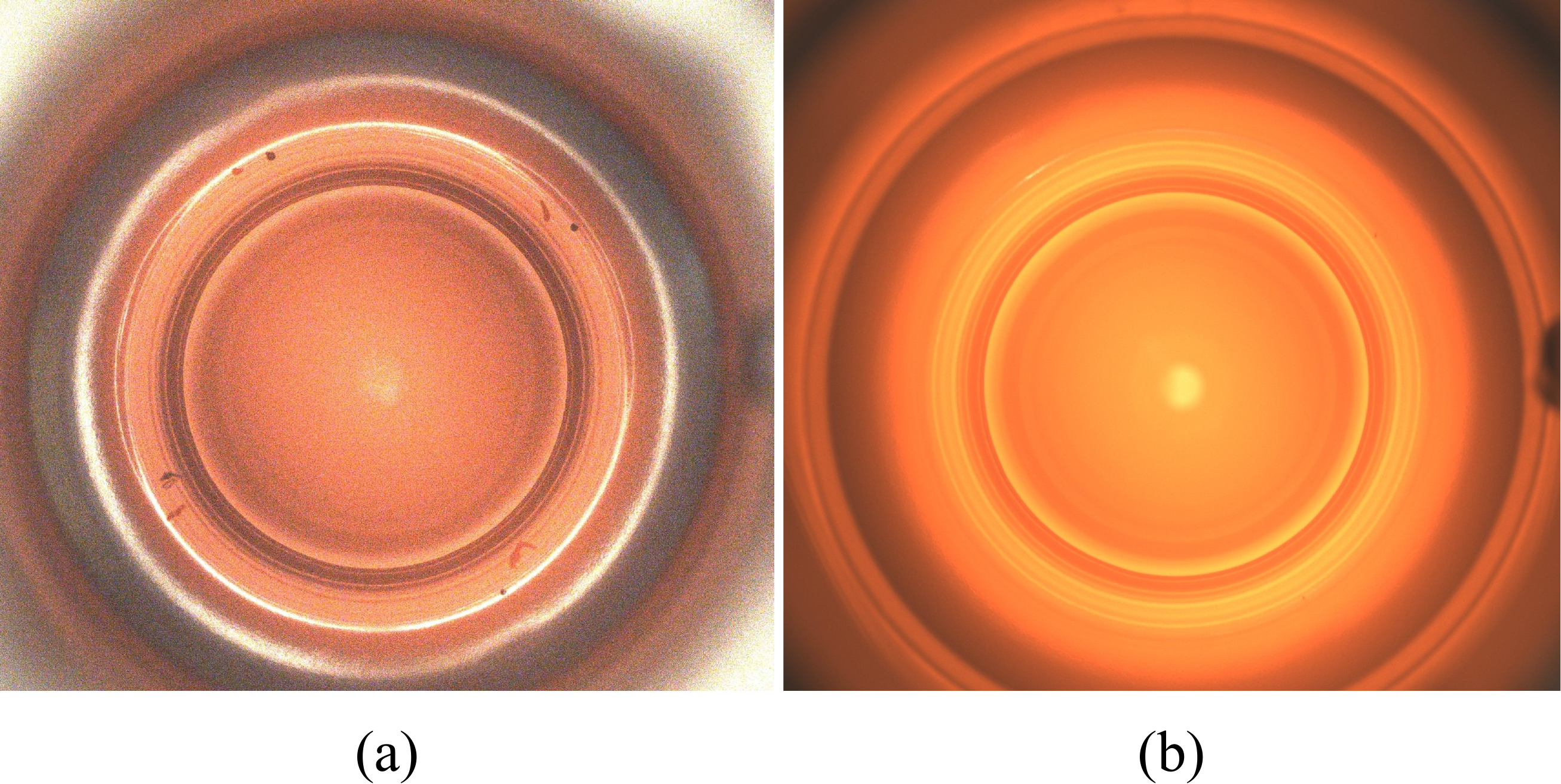}
	\caption{Glow discharge ignited in the end cell contaminated with permanent marker dots. (a) initial state, (b) final state after 19 hours of plasma processing.}
    \label{fig:photomarker}

    \centering
    \includegraphics[width=0.75\columnwidth]{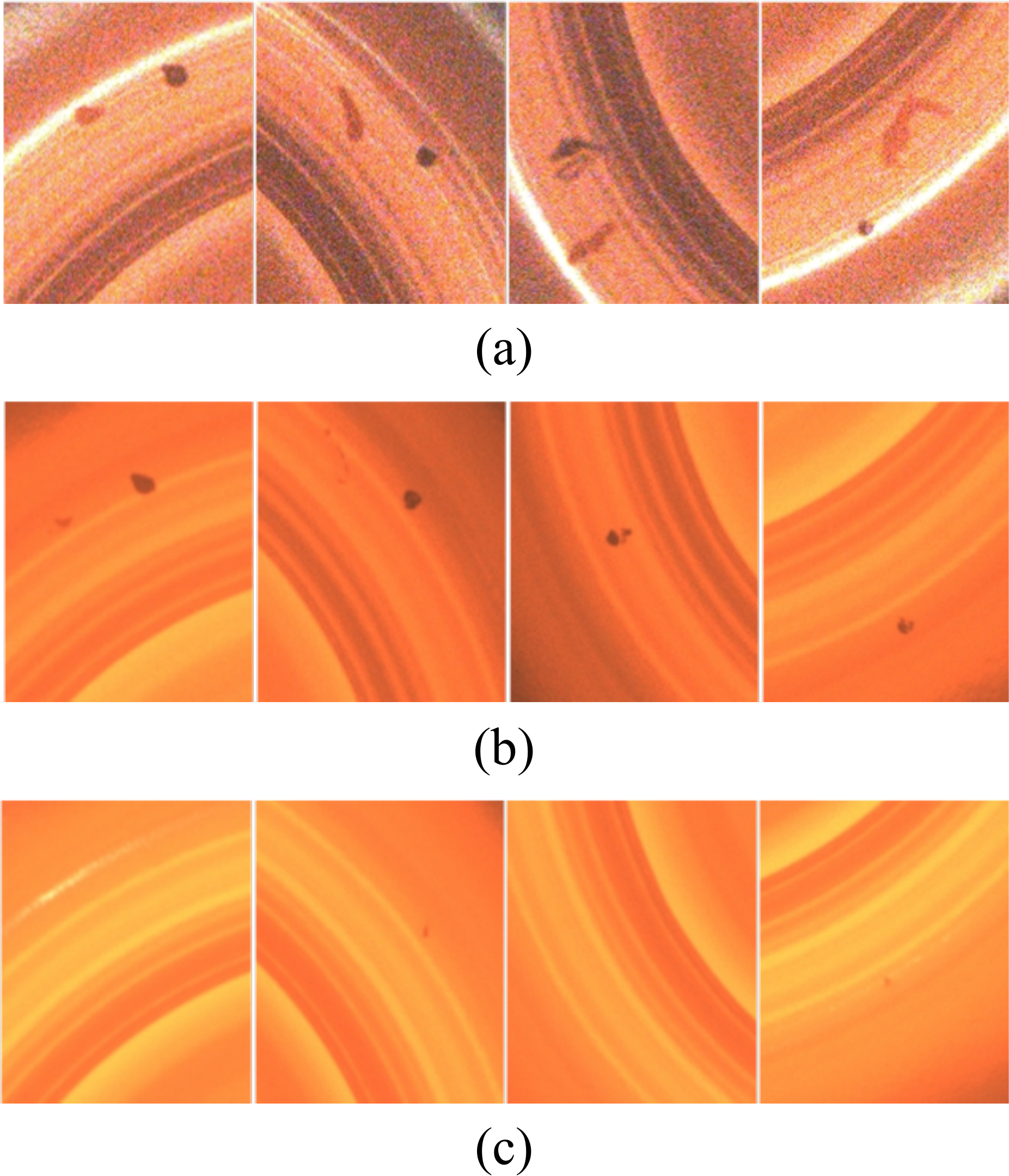}
	\caption{Close up of the contamination showed in Fig. \ref{fig:photomarker}. Initial state of the contamination is shown in (a), final state in (c); (b) shows the progress after 5 hours of plasma processing.}
	\label{fig:photomarkerzoom}
\end{figure}

After the plasma ignition studies on \SI{1.3}{\giga\hertz} cavities \cite{berrutti2019plasma}, the first plasma processing test was performed on a 9-cell cavity assembled with viewports on the beam tubes. A permanent marker was used to introduce a carbon-based contamination on the iris of one end cell. Permanent marker ink is composed of hydrocarbon chains, as shown by the EDS (Energy Dispersive X-ray spectroscopy) analysis in Fig. \ref{fig:ink_EDS}, and it was previously used for plasma processing studies at SNS \cite{doleans2013plasma, tyagi2014plasma}. Two permanent markers (black and red ink) were used to draw 8 dots on the cavity iris. We applied plasma processing to the contaminated cell for \SI{19}{\hour}: Fig. \ref{fig:photomarker} shows the initial and final state of the cavity, while Fig. \ref{fig:photomarkerzoom} offers a close up of the initial, intermediate and final state.

With this test, we confirmed that plasma processing is able to remove a \ch{C}-based contamination from the cavity iris, the area most problematic for FE. The experiment has also allowed us to develop a first plasma recipe in terms of duration of the process, \ch{O2} percentage, pressure, and plasma density. This set of parameters was later used to apply plasma processing on multiple LCLS-II cavities, both single cell and 9-cell cavities.

\section{\label{sec:RFtests} Plasma processing and RF tests on \SI{1.3}{\giga\hertz} Cavities}
Plasma processing was applied to LCLS-II cavities, both single cell and 9-cells, and the effectiveness of plasma cleaning was measured in terms of $\mathrm{Q_0}$ versus $\mathrm{E_{acc}}$ and radiation versus $\mathrm{E_{acc}}$ curves. All the cavities were cold tested before and after plasma processing in order to compare their performance.\\
The first test was carried out on a clean nitrogen doped cavity, the following two on naturally field emitting 9-cell N-doped cavities. The subsequent studies were conducted on cavities artificially contaminated in order to investigate the efficacy of plasma processing under different circumstances. Table \ref{tab:summary} summarizes the tests conducted on LCLS-II \SI{1.3}{\giga\hertz} cavities.

\begin{table*}%The best place to locate the table environment is directly after its first reference in text
    \caption{\label{tab:summary}Summary of the \SI{1.3}{\giga\hertz} single cell and 9-cell TESLA-shaped cavities cold tested for plasma processing studies. The terminology 'n/m' for the nitrogen doping is to be interpreted as $n$ minutes in 25\,mTorr of nitrogen and $m$ minutes of annealing in vacuum.}
\begin{ruledtabular}
\begin{tabular}{llll}
Cavity & Surface Treatment & Contamination & Test Scope \\ \hline
Single cell & '2/6' N-doped & & Plasma Processing Effect on N-doping \\
9-cell & '3/60' N-doped & Natural FE & Removal of Natural FE \\
9-cell & '3/60' N-doped & Natural FE & Removal of Natural FE \\
Single cell & '2/6' N-doped & Aquadag\textsuperscript{\textregistered} & Removal of C-contamination \\ 
9-cell & EP & Vacuum Failure Simulated Inside Cleanroom & FE Mitigation \\
Single cell & '2/6' N-doped & Vacuum Failure Simulated Outside Cleanroom & FE Mitigation \\
9-cell & '2/6' N-doped & Vacuum Failure Simulated Outside Cleanroom & FE Mitigation \\
9-cell & '2/6' N-doped & Vacuum Failure Simulated Outside Cleanroom & FE Mitigation \\
\end{tabular}
\end{ruledtabular}
\end{table*}

All the RF cold tests were done in the vertical test stand (VTS) facility at FNAL, following the measurement method explained in \cite{melnychuk2014error}. The cryogenic dewars are equipped with two radiation detectors \cite{krueger2002chipmunk} positioned on the top and on the bottom of the dewar where the cavity is placed for the cold test. In order to reliably compare the results of the RF tests done before and after plasma processing, we have attempted to test each cavity always in the same VTS dewar and, when possible, in the same position inside the dewar. In some cases it was not possible to test the cavity in the same dewar and it is indicated in the text.

All the plots in this paper use the following symbols: solid symbols for $\mathrm{Q_0}$ versus $\mathrm{E_{acc}}$ curves, empty and half filled symbols for the radiation versus $\mathrm{E_{acc}}$ curves, with empty symbols used for the radiation detector located on top of the cryogenic dewar, vertically half filled symbols for the bottom radiation detector.

\subsection{\label{sec:Ndoped} Baseline test on N-doped Cavity}
N-doping is a surface treatment developed at FNAL that has allowed to increase the cavity $\mathrm{Q_0}$ by a factor of 3 \cite{grassellino2013nitrogen, grassellino2015n}. The recipe used to N-dope LCLS-II cavities is called '2/6', which consists of baking the cavity in vacuum (p $<$1E-6 Torr) at \SI{800}{\celsius}, once the temperature is stable nitrogen is injected in the furnace at a pressure of 25$\,$mTorr for \SI{2}{\minute}. Afterwards the vacuum is restored and the cavity undergoes \SI{6}{\minute} of annealing.  After the doping, \SI{5}{\micro\meter} are removed from the inner cavity surface through electropolishing (EP) in order to eliminate possible nitrides. After \SI{5}{\micro\meter} of EP, only the interstitial nitrogen remains in the niobium with a concentration in the order of 100 ppm. \ch{N_2} atoms are absorbed as interstitial impurities in the niobium lattice and cause a reduction of the Mattis-Bardeen surface resistance $\mathrm{R_{BCS}}$ with the accelerating field \cite{martinello2016effect}.

\begin{figure}
    \centering
    \includegraphics[width=1\columnwidth]{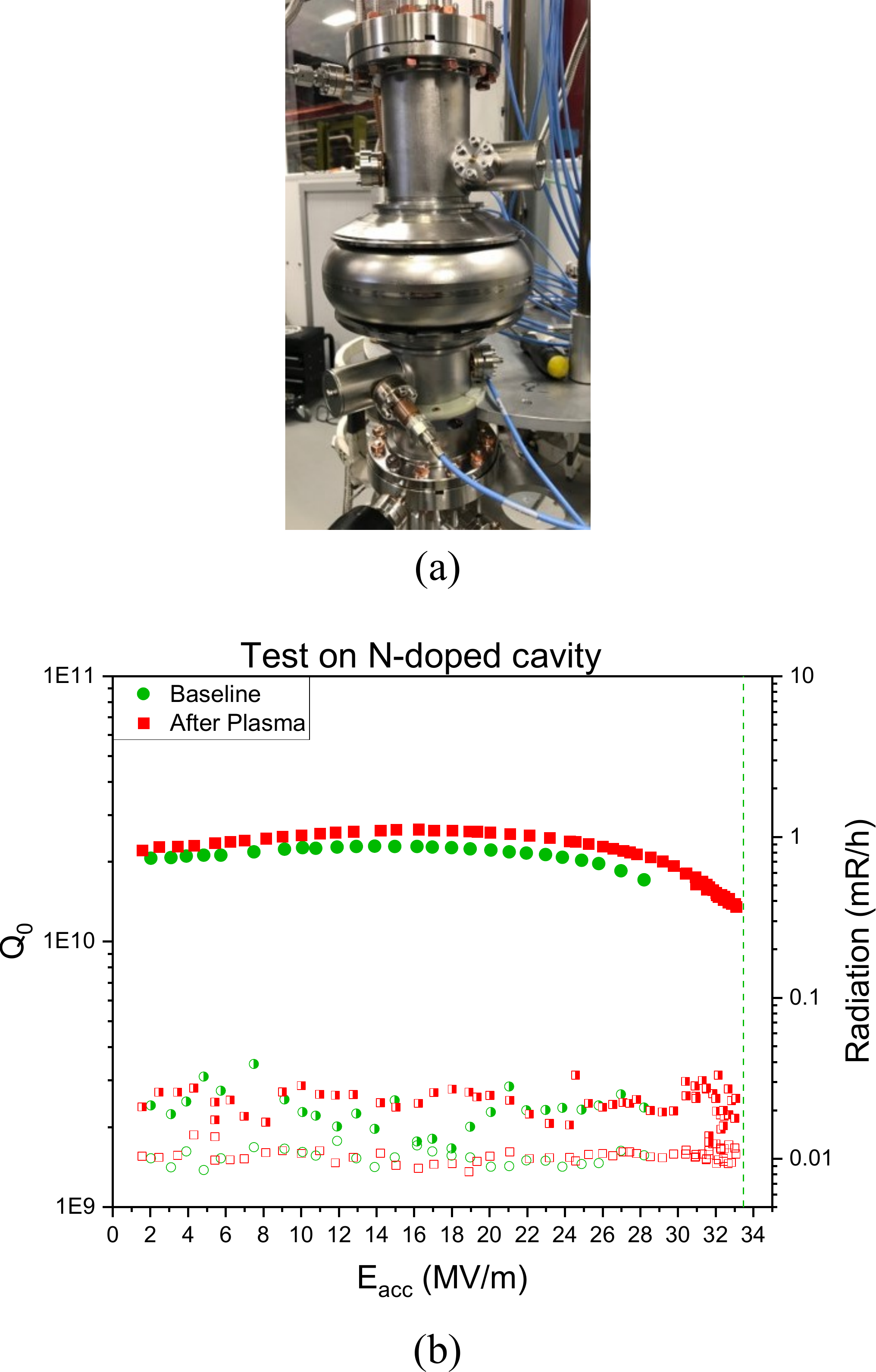}
    \caption{In (a) single cell cavity with HOM couplers. Figure (b) shows the results of the RF tests measured at \SI{2}{\kelvin} before (green) and after (red) plasma processing. The baseline RF test has been intentionally limited at 28\,MV/m at \SI{2}{\kelvin} to avoid quench. The quench field was reached at $\mathrm{E_{acc}}=33.5$\,MV/m during the \SI{1.4}{\kelvin} baseline test and the vertical dotted line indicates the quench field value. As explained in section \ref{sec:RFtests} solid symbols are used to plot the $\mathrm{Q_0}$ vs $\mathrm{E_{acc}}$ curve, empty symbols for the radiation detected from the top detector, half filled symbols for the bottom radiation detector.}
    \label{fig:Ndoping_test}
\end{figure}

Since all \SI{1.3}{\giga\hertz} LCLS-II cavites are nitrogen doped \cite{crawford2014joint, gonnella2018industrialization}, we have started our studies applying plasma cleaning to a N-doped single cell cavity in order to understand if this processing can affect the surface treatment.\\
The single cell used for the test was built by welding together two end-cells of a 9-cell cavity. The result is a single cell cavity with HOMs couplers on the beamtubes (sew panel (a) in Fig. \ref{fig:Ndoping_test}). This characteristic makes it suitable for plasma processing, as it allows to ignite the glow discharge using the HOMs and, with only one cell to process, drastically reduces the duration of the cleaning.

The results of the RF cold tests obtained on the single cell before and after plasma processing are shown in panel (b) Fig. \ref{fig:Ndoping_test}. We intentionally stopped the baseline test before quench in order to measure the cavity $\mathrm{Q_0}$ at \SI{1.4}{\kelvin}, where it reached a quench field equal to 33.5\,MV/m. After the first vertical test, the cavity was connected to the vacuum/gas and RF system used for plasma cleaning and processed for \SI{16}{\hour} with \ch{Ne}$/$\ch{O2} plasma.\\
Comparing the RF tests measured before and after plasma processing, it is clear that plasma cleaning does not negatively affect the performance of the nitrogen doped single cell; on the contrary, it preserves the high quality factor and quench field characteristic of N-doped cavities.

\subsection{\label{sec:naturalFE} Naturally Field Emitting N-doped Cavities}
\begin{figure}[htb!]
    \centering
    \includegraphics[width=1\columnwidth]{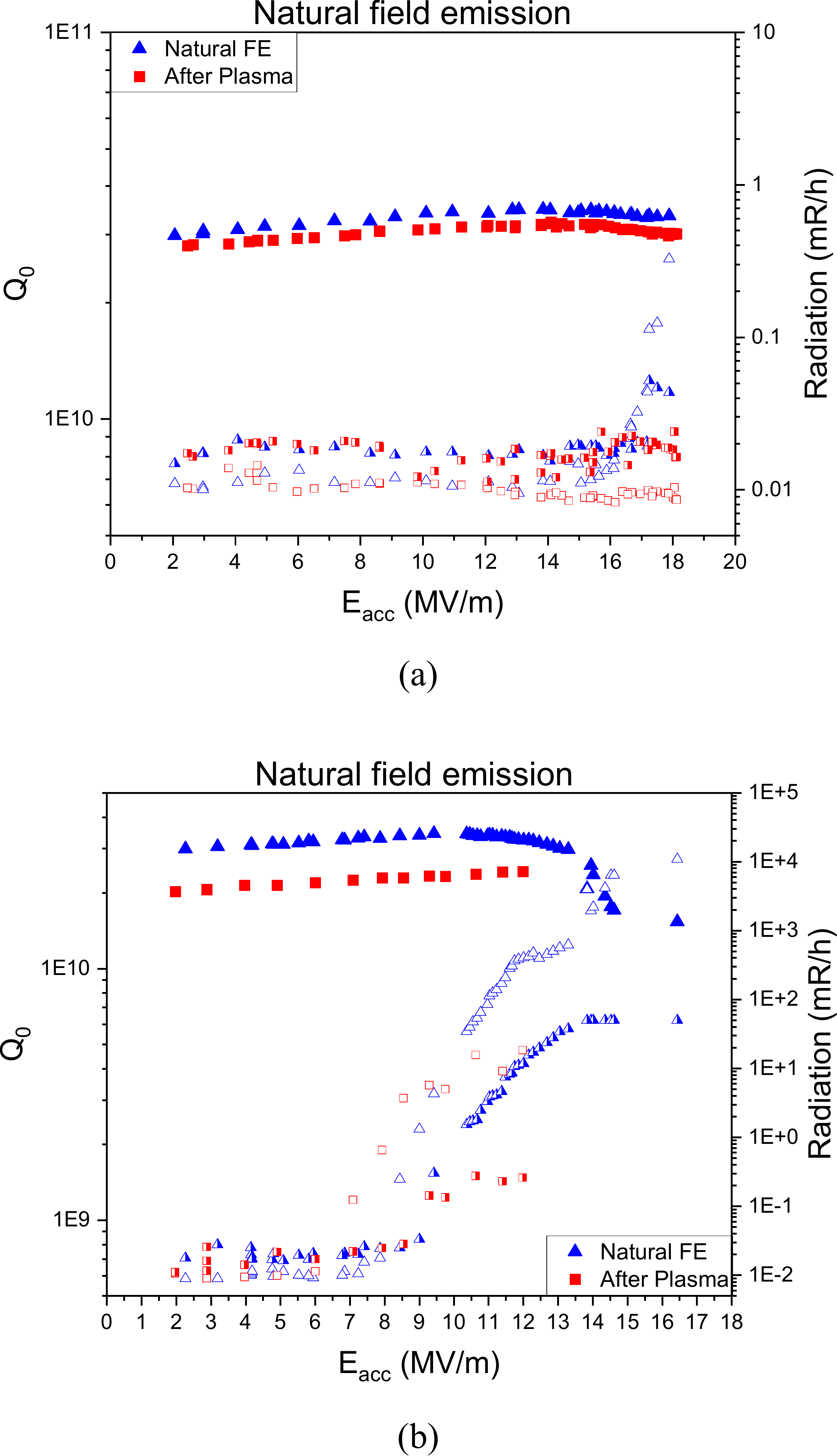}
    \caption[RF tests of Natural FE]{Figure (a) shows the results of the first naturally field emitting cavity: in blue $\mathrm{Q_0}$ and radiation level vs $\mathrm{E_{acc}}$ performed before plasma processing. The X-ray onset is at 16\,MV/m before plasma processing; after plasma processing the FE was completely removed: the RF test measured after the treatment shows no X-rays. Figure (b) shows the RF tests of the second cavity. The test was stopped before quench due to intense X-rays. The red curves show the performance after plasma processing; no increase in performance has been registered after the cleaning.}
   \label{fig:naturalFE_tests}
\end{figure}

We used two cavities with natural field emission to test the efficacy of plasma processing on FE with unknown source, not caused by an artificial contamination. The two 9-cell cavities showed X-rays during the first vertical tests performed at FNAL. Both cavities were assembled with a second valve to allow the flow of gas for plasma processing and RF tested again after valve assembly. Afterwards the two cavities were plasma processed and cold tested again.

The top plot in figure \ref{fig:naturalFE_tests} shows the results of the cold tests of the first 9-cell; the curves registered before plasma show that the cavity quenched at $\mathrm{E_{acc}}\,=\,18.5\,$MV/m with X-rays onset at 16$\,$MV/m. After plasma processing, the cavity reached $\mathrm{E_{acc}}\,=\,18\,$MV/m showing no X-rays. In this case plasma processing completely removed field emission. The fact that no change in quench field has been registered suggests that is a hard quench, not due to FE.

The bottom plot in figure \ref{fig:naturalFE_tests} shows the performance of the second cavity: the radiation onset before plasma processing was registered at 7\,MV/m; the test was stopped before the quench field due to intense radiation levels (1.1E4\,mR/h at 16.5\,MV/m) and final FE onset was measured at 7.8\,MV/m. The cold test conducted after plasma processing showed the X-ray onset decreased to 7\,MV/m and the quality factor degraded. Also in this case, the RF test was interrupted due to intense FE. The $\mathrm{Q_0}$ degradation could be due to higher ambient magnetic field during cooldown, which results in increased trapped magnetic flux in the cavity. The two RF tests measured on this cavity were carried out in different cryogenic dewar, this could also explain the difference in trapped flux.

The fact that plasma processing was effective on one cavity with natural field emission but not on the second indicates that the FE may originate from different sources. The evidence suggests that in the first cavity the FE was caused by a hydrocarbon contamination and that plasma processing was effective in removing it thanks to the reactive oxygen, present in the glow discharge, that binds with the \ch{H_xC_y} and creates volatile byproducts. In the second cavity instead the field emission may not be caused by a hydrocarbons but it could be due to surface defects or metal flakes on the cavity surface. If that is the case, it is expected that the plasma cleaning would have little effect in mitigating FE since no volatile byproducts are generated in the reaction.

\subsubsection{Residual Gas Analyzer spectrum}
\begin{figure}
   \includegraphics[width=1\columnwidth]{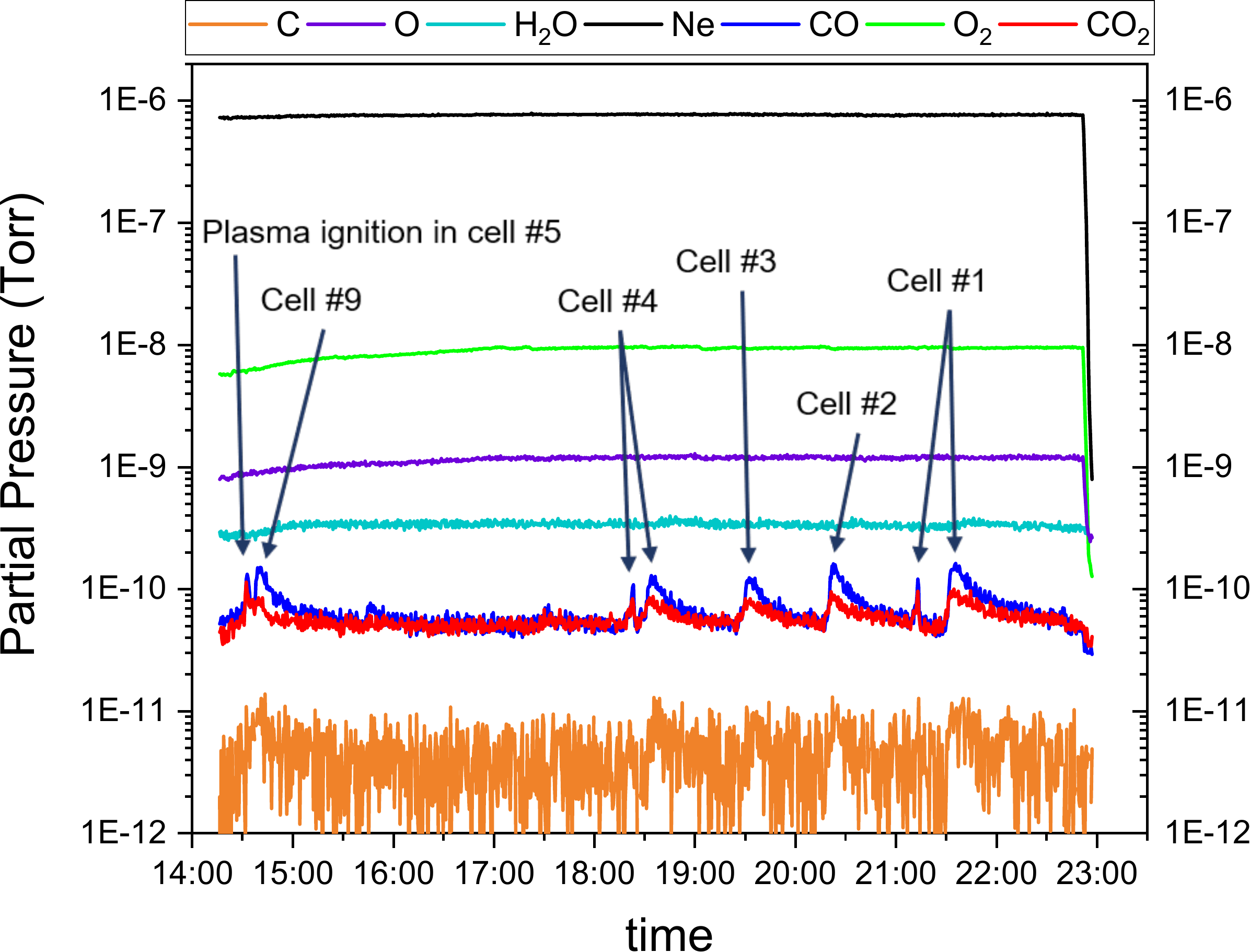}
   \caption{\label{fig:RGAplot}Example of RGA spectrum acquired during the first day of plasma processing on a 9-cell cavity. Cell $\#$4 and cell $\#$1 exhibit a double peak due to the plasma being accidentally turned off while reaching the desired plasma density.}
\end{figure}

A Residual Gas Analyzer is used to monitor the composition of the gas pumped out of the cavities for the duration of the entire plasma cleaning.
Figure \ref{fig:RGAplot} shows the RGA data acquired during the first round of plasma processing applied to the first naturally field emitting cavity and it represents a typical example of the RGA spectrum registered during plasma processing.
During the first round of plasma cleaning, the spectrum often shows peaks in \ch{C}, \ch{CO}, \ch{CO2} in correspondence with the moment when the plasma is ignited in, or transferred to, a new cell. The increase in the \ch{C}-related signals shows that the oxygen is reacting with the \ch{H_xC_y} on the cavity surface.

The RGA can record masses from 1 to \SI{300}{\amu}, however in this plot are shown only the elements of interest.
In figure \ref{fig:RGAplot} the \ch{C}-related peaks are clearly visible in correspondence with the plasma being initially ignited in cell $\#$ 5, then transferred to, and tuned, in cell $\#$ 9, and transferred and tuned in cell $\#$ 4, up to cell $\#$ 1. Cells from $\#$ 8 to $\#$ 5 don't show prominent peaks in the \ch{C}-related signals. This could be due to the fact that, at the beginning of the procedure, the glow discharge is ignited in the central cell (first peak on the left in figure \ref{fig:RGAplot}) and then moved from cell $\#$ 5 up to cell $\#$ 9, going through all the intermediate cells.

\subsection{\label{sec:Aquadag} Carbon-based contamination}
\begin{figure}[hbt!]
    \centering
    \includegraphics[width=0.75\columnwidth]{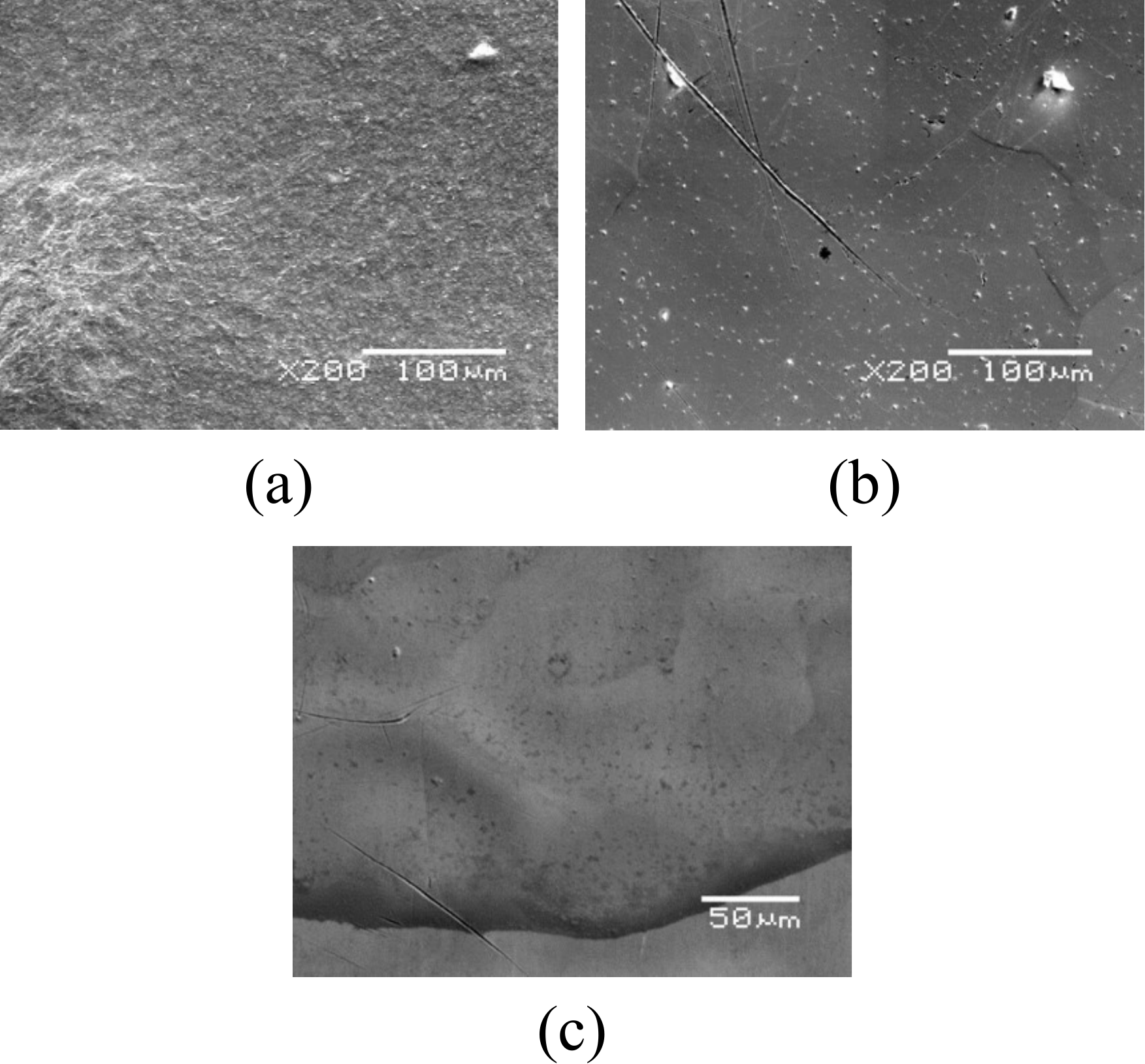}
    \caption[SEM images of Aquadag]{Scanning Electron Microscope images of pure (a) and diluted Aquadag\textsuperscript{\textregistered} on Nb substrate. The dilution factor has been calculated as the ratio between \ch{H2O} mass and Aquadag mass. Figure (b) and (c) show respectively Aquadag diluted by a factor of 100 and 2E4.}
	\label{fig:aquadagsem}

   \centering
   \includegraphics*[width=1\columnwidth]{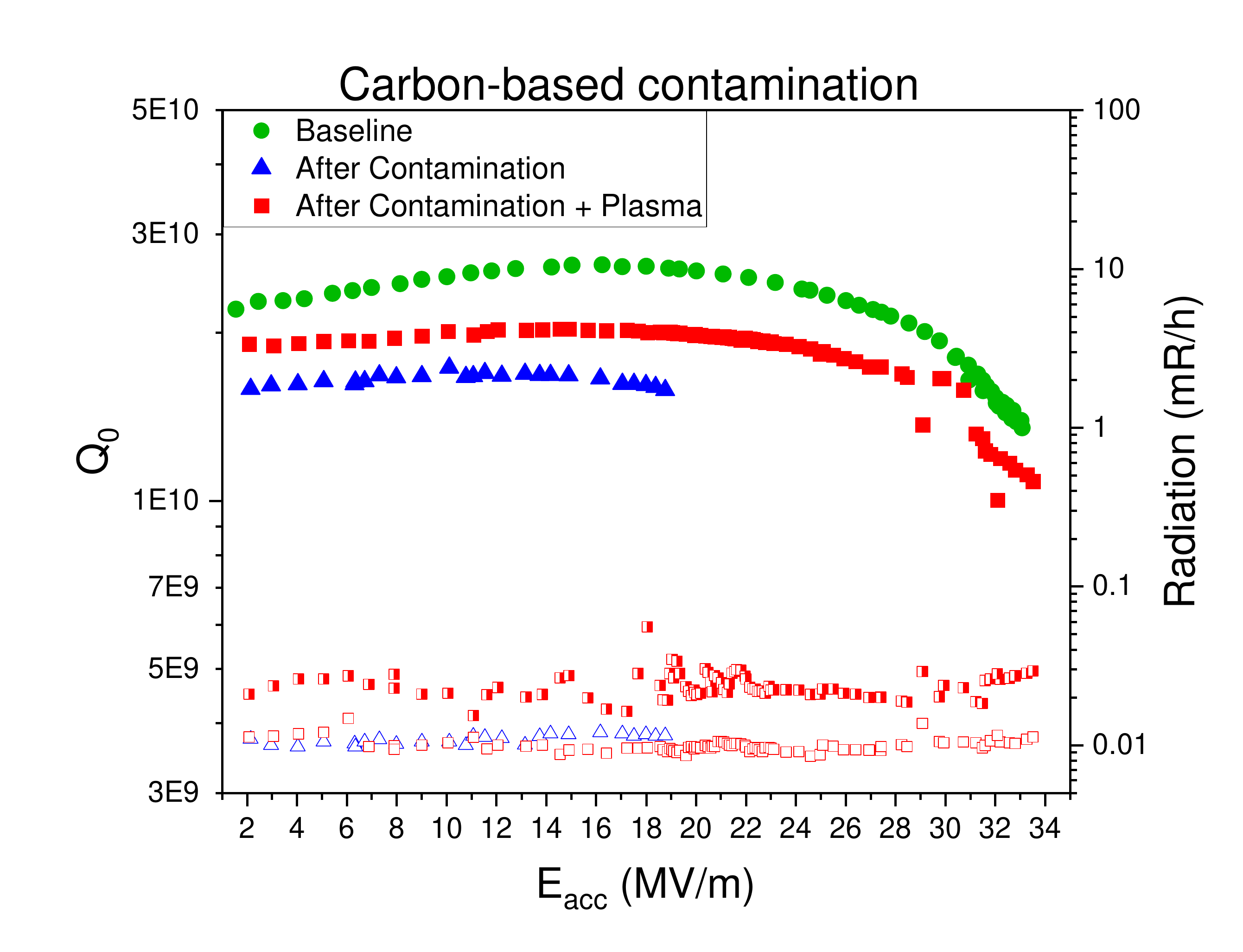}
   \caption{Quality factor versus accelerating field curves measured on the contaminated cavity and after plasma cleaning. The contaminated cavity (blue triangle) shows a degradation in quench field and $\mathrm{Q_0}$. After plasma processing it shows an increase in quality factor ($\mathrm{Q_0}=$\,2E10 at 16.4\,MV/m) and a complete recovery in $\mathrm{E_{acc}}$ (quench field is 33.5\,MV/m). In green is shown the baseline performance of the cavity before contamination.}
   \label{fig:aquadag_test}
\end{figure}

We contaminated the single cell cavity, previously used for the N-doping test, using a carbon-based paint in order to study the effectiveness of plasma processing on an artificial carbon contamination. Aquadag\textsuperscript{\textregistered} \cite{Aquadag}, a conductive paint made of graphite and ultra pure water, was used to contaminate the cavity. A small drop of highly diluted paint was deposited on the cavity iris. Figure \ref{fig:aquadagsem} shows images of pure and diluted Aquadag acquired with the Scanning Electron Microscope (SEM). The Aquadag used to contaminate the single cell iris was diluted by a factor 2E4, where the dilution was calculated as the ratio between the \ch{H_2O} and the Aquadag mass.

For this study it was not possible to cold test the cavity in the same cryogenic dewar before and after plasma processing.

Figure \ref{fig:aquadag_test} shows the results of the RF tests. The contaminated single cell was plasma processed for \SI{17}{\hour} between the two cold tests. In blue is the curve measured on the contaminated cavity; comparing it with the baseline test (here in green, same curve shown in red in Fig. \ref{fig:Ndoping_test}), it can be seen that the cavity shows a degradation in the quality factor and quench field: $Q_0\,=$\,1.7E10 at $\mathrm{E_{acc}}\,=$\,16.2\,MV/m, quench field is registered at 18.8\,MV/m. The radiation detector positioned at the bottom of the cryogenic dewar was not working correctly during this cold test, however the top radiation detector was connected and no X-rays were registered. After \SI{17}{\hour} of plasma processing, the cavity exhibits an increase in quality factor ($\mathrm{Q_0}\,=$\,2E10 at $\mathrm{E_{acc}}\,=$\,16.4\,MV/m). Plasma processing increased the quench field by almost 15\,MV/m, restoring the initial quench field at $\mathrm{E_{acc}}\,=$\,33.5\,MV/m.

\subsection{\label{sec:VacuumFailure} Vacuum Failure Experiments}
A possible cause of cavity contamination is a vacuum leak or a complete vacuum loss.
Multiple experiments were conducted on cavities exposed to air in order to understand if plasma processing can be effective in mitigating field emission in these scenarios. The tests were carried out under different conditions, on both 9-cell and single cell cavities. We refer to these tests as vacuum failure experiments (or simulations).

\subsubsection{\label{sec:VacuumFailure_InCR} Vacuum Failure Experiment Inside the Cleanroom}
We conducted the first test inside a cleanroom environment, in order to introduce a controlled amount of particulate. High pressure rinsing (HPR) was used to clean the cavity and, after drying, it was slowly evacuated to high vacuum. To simulate the vacuum failure, the mini right angle valve (RAV) was opened while the cavity was in a class 100 cleanroom. The cavity quickly reached atmospheric pressure and, after sitting at this pressure for a few minutes, it was slowly evacuated to reach a pressure in the low E-6\,-\,high E-7\,Torr range.

Plasma processing was applied twice to this cavity, each time using the standard parameters and duration (approximately \SI{1}{\hour} \SI{40}{\minute} per cell). After each plasma processing, the cavity was RF tested at \SI{2}{\kelvin}. The RF tests on the contaminated cavity and after the second plasma processing were conducted in the same cryogenic dewar, while the cold test after the first plasma cleaning was carried out in a different dewar.

\begin{figure*}
    \centering
    \includegraphics[width=1\textwidth]{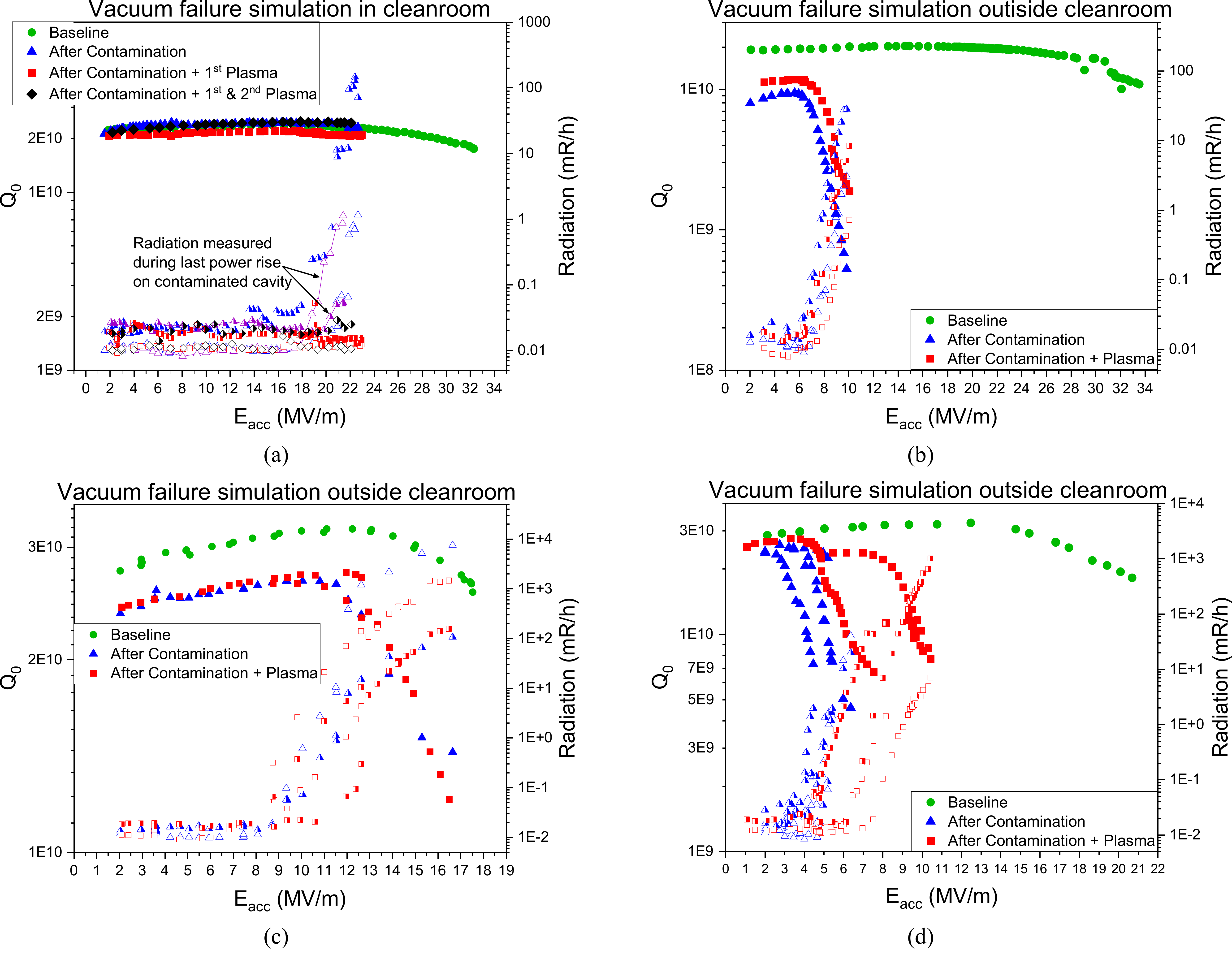}
    \caption[Vacuum failures simulated inside and outside cleanroom]{Vacuum failure experiments simulated on four cavities. Figure (a) shows the curves of the 9-cell cavity vented in cleanroom; the results of the cold test perfored after the cavity was contaminated is plotted in blue; the final radiation curves were acquired during this test at \SI{1.4}{\kelvin}, they are plotted with a solid line and purple triangles (empty triangles for the top radiation detector, half filled for the bottom detector). This cavity was plasma processed twice: the cold test conducted after the first round of plasma cleaning is plotted red, in black is the cold test performed after the second plasma round. Figure (b) contains the RF tests of the single cell cavity vented outside the cleanroom through the RAV (right angle valve), both RF tests (after the contamination and the subsequent after plasma test) were stopped before quench at 10\,MV/m; Fig. (c) shows the results of the 9-cell cavity quickly vented through the mini RAV; plot (d) contains the curves of the 9-cell cavity slowly vented through the mini RAV.\\
   For comparison all the plots also contain the baseline curves of the cavities, measured before they were exposed to the vacuum failure simulations.}
   \label{fig:vacuumfailures}
\end{figure*}

Panel (a) of figure \ref{fig:vacuumfailures} summarizes the performance of the cavity during the cold tests. In blue the curves registered before plasma processing (on the contaminated cavity): the 9-cell reached a first quench at 7.5\,MV/m, it was then possible to increase the power and measure the $\mathrm{Q_0}$ versus $\mathrm{E_{acc}}$ curve up to 23\,MV/m (with intermediate quenches at 20.5 and 22\,MV/m), when the cavity reached the final quench. The X-ray onset was registered at 18.5\,MV/m for the bottom radiation detector, at 20\,MV/m for the top detector. The cavity was tested also at \SI{1.4}{\kelvin}, and the purple curves show the radiation level registered during the final power rise (intentionally stopped before quench): the maximum radiation level switched from the bottom to the top detector, the onset was registered at 16\,MV/m and 19\,MV/m respectively for bottom and top detector. This, along with the change in radiation intensity, suggests that some field emission processing has occurred during the RF test, however the cavity continued showing significant FE.
The RF test measured after the first plasma processing is shown in red: the cavity quenched at 23\,MV/m and no X-ray activity was registered during the test, indicating that the field emission was completely removed. The quality factor instead showed degradation: from 2.4E10 at 16.3\,MV/m registered before plasma to 2.2E10 at 16.4\,MV/m after plasma processing.\\
We applied a second round of plasma processing to the 9-cell cavity to investigate if the quality factor degradation was caused by the plasma treatment or if it was due to different amount of trapped flux since the two RF tests (after the contamination and after the first plasma processing) were measured in different dewars. After an additional \SI{2}{\hour} of plasma cleaning per cell, we cold tested the cavity. The curves are plotted in black in panel (a) figure \ref{fig:vacuumfailures}: the test confirms that the quality factor is actually preserved ($\mathrm{Q_0}$=2.5E10 at 16.4\,MV/m) and both radiation detectors show no X-ray activity, confirming that the field emission was eliminated.

\subsubsection{\label{sec:VacuumFailure_OutCR} Vacuum Failure Experiments Outside the Cleanroom}
Following the experiment in cleanroom, additional vacuum failures were simulated outside the cleanroom. The procedure was repeated with small variations on three cavities: one single cell and two 9-cells.
After the contamination, the single cell cavity was plasma processed for \SI{20}{\hour}, and the 9-cell cavities for approximately \SI{1}{\hour} \SI{40}{\minute} per cell.
After the venting, the cavities were evacuated to the low E-6\,-\,high E-7\,Torr range; RF tests were performed before and after plasma processing. The plots in Fig. \ref{fig:vacuumfailures} show the $\mathrm{Q_0}$ and radiation versus $\mathrm{E_{acc}}$ curves.

The single cell cavity was quickly vented through the mini RAV from high vacuum to atmospheric pressure. 
Panel (b) in figure \ref{fig:vacuumfailures} contains the RF tests of the single cell cavity. The green curve shows the baseline before the vacuum failure experiment (red curve in Fig. \ref{fig:aquadag_test}, $\mathrm{Q_0}$=1.9E10 at 5\,MV/m). The contaminated cavity (blue curves) exhibits a degradation in quality factor ($\mathrm{Q_0}$=9.4E9 at 5\,MV/m) and radiation onset at 6.2\,MV/m. The test was limited before quench at 10\,MV/m by the available RF power. Due to the severity of FE registered during the first RF test it was decided to process the cavity for a time longer than usual (\SI{20}{\hour}). The cold test conducted after plasma processing shows a moderate increase in quality factor ($\mathrm{Q_0}$=1.2E10 at 5.1\,MV/m) and decrease in radiation (8.4\,mR/h at 10\,MV/m versus the 28\,mR/h at 9.8\,MV/m registered before plasma) with FE onset at 7\,MV/m. \\

Panel (c) in figure \ref{fig:vacuumfailures} shows the results of the 9-cell cavity quickly vented through the mini RAV. In green is plotted the baseline performance of the cavity before venting ($\mathrm{Q_0}$=3.1E10 at 8.8\,MV/m); in blue are the curves measured on the cavity after the contamination: quality factor degradation ($\mathrm{Q_0}$=2.6E10 at 8.7\,MV/m) and FE onset at 8.7\,MV/m, reaching final quench at 17\,MV/m with 1E4\,mR/h; the FE onset after quench was registered at 10.3\,MV/m. The after plasma curves are shown in red: FE onset at 8.8\,MV/m, with less severe slope in $\mathrm{Q_0}$ versus $\mathrm{E_{acc}}$ curve until 12.6\,MV/m; at this point an increase in radiation was registered along with a $\mathrm{Q_0}$ drop. The $\mathrm{Q_0}$ versus $\mathrm{E_{acc}}$ curve was measured again showing overlap with the test measured on the contaminated cavity in terms of quench field (17\,MV/m), quality factor degradation and slope due to FE.

We simulated the final vacuum failure experiment by slowly opening the mini RAV on a 9-cell cavity, venting the cavity over a \SI{18}{\minute} time interval. The results of the RF tests are shown in panel (d) Fig. \ref{fig:vacuumfailures}. The green curve shows the baseline performance; in blue are plotted the curves of the contaminated cavity (before plasma): the cavity shows intense FE as evident from the slope in quality factor. A first quench was reached below 3\,MV/m, we then increased the accelerating field: at 4\,MV/m FE started and between 4.5-5\,MV/m there was a switch in quality factor (from 7.3E9 to 1.75E10). The sudden jump in $\mathrm{Q_0}$ suggests that a field emitter was processed by the RF, eliminating the radiation and energy dissipation that it was causing. After this event the quality factor followed a new curve that started bending above 4\,MV/m while the radiation increased. Another $\mathrm{Q_0}$ switch occurred at 6.4\,MV/m, indicating that a new field emitter was likely processed. The test was stopped at this point due to intense FE ($>$1\,R/h) and to avoid risking damaging the cavity surface through RF processing of field emitters; the final FE onset was registered at 5.2\,MV/m. In red is reported the RF test measured after plasma processing: the $\mathrm{Q_0}$ curve overlaps with the final before plasma (blue) curve. Also in this case a $\mathrm{Q_0}$ switch was observed at 7.5\,MV/m, indicating RF processing of the field emitter. The test was stopped at 10.4\,MV/m. The partial performance recovery observed in this cavity is to be attributed to RF processing occurred during the \SI{2}{\kelvin} vertical tests, not to plasma processing.

\section{\label{sec:conclusions} Conclusions}
Using the newly developed technique of plasma ignition with HOMs, it was demonstrated that plasma processing successfully interacts with the cavity iris and removes carbon-based contamination.\\
It was also proved that plasma processing does not affect negatively the performance of nitrogen doped cavities, on the contrary it preserves their high quality factor and quench field.
Plasma cleaning was applied to multiple LCLS-II cavities with natural field emission or artificially contaminated. The comparison between the RF tests conducted before and after plasma cleaning showed an increase in performance in the carbon contaminated single cell, in one out of two naturally field emitting cavities and in one 9-cell cavity exposed to vacuum failure simulation inside the cleanroom. A second cavity with natural field emission was processed but still showed X-rays activity after the plasma, suggesting that the source of FE may not be carbon-related in this case, but due to metal flakes or surface defects. The three cavities used for vacuum failure simulation outside the cleanroom have shown little or no improvement attributable to plasma processing.

Plasma processing has shown positive results in cavities with field emission onset registered at high field level (above 16\,MV/m), while it has not been able to reduce FE in cases when the onset started at low fields. This can be related with the source of FE: plasma processing, applied with the current recipe and parameters, is effective on hydrocarbon contamination and not on metal flakes, which are the most plausible cause of FE at low fields.

In order to better understand what is the nature of the field emitters in the cavities exposed to vacuum failure simulations, it was decided to collect and analyze the particles introduced in the cavities and not removed with plasma cleaning. Preliminary results from the analysis of particles collected from the single cell cavity suggest that metal flakes were introduced into the cavity during the vacuum failure experiment performed outside the cleanroom. The SEM/EDS analysis of the particles collected from the vented cavities is currently ongoing and will be the subject of future publication.

We intend to apply plasma processing to more LCLS-II 9-cell cavities and cold test them (before and after) in order to acquire a greater statistics, focusing in particular on cavities that exhibit field emission of unknown source (natural FE) during the RF tests.

\begin{acknowledgments}
This work was supported by the U.S. Department of Energy (DOE), Office of Science,
Basic Energy Sciences (BES). Fermilab is operated by Fermi Research Alliance, LLC under Contract No. DE-AC02-07CH11359 with the U.S. Department of Energy.

We thank LCLS-II-HE for providing the N-doped \SI{1.3}{\giga\hertz} 9-cell cavities used in this study for plasma processing and cold tests.
\end{acknowledgments}

\bibliography{BGiaccone_Plasma_2020}% Produces the bibliography via BibTeX.

%aipnum4-2.bst 2019-01-14 (MD) hand-edited version of apsrev4-1.bst
\begin{thebibliography}{36}%
\makeatletter
\providecommand \@ifxundefined [1]{%
 \@ifx{#1\undefined}
}%
\providecommand \@ifnum [1]{%
 \ifnum #1\expandafter \@firstoftwo
 \else \expandafter \@secondoftwo
 \fi
}%
\providecommand \@ifx [1]{%
 \ifx #1\expandafter \@firstoftwo
 \else \expandafter \@secondoftwo
 \fi
}%
\providecommand \natexlab [1]{#1}%
\providecommand \enquote  [1]{``#1''}%
\providecommand \bibnamefont  [1]{#1}%
\providecommand \bibfnamefont [1]{#1}%
\providecommand \citenamefont [1]{#1}%
\providecommand \href@noop [0]{\@secondoftwo}%
\providecommand \href [0]{\begingroup \@sanitize@url \@href}%
\providecommand \@href[1]{\@@startlink{#1}\@@href}%
\providecommand \@@href[1]{\endgroup#1\@@endlink}%
\providecommand \@sanitize@url [0]{\catcode `\\12\catcode `\$12\catcode
  `\&12\catcode `\#12\catcode `\^12\catcode `\_12\catcode `\%12\relax}%
\providecommand \@@startlink[1]{}%
\providecommand \@@endlink[0]{}%
\providecommand \url  [0]{\begingroup\@sanitize@url \@url }%
\providecommand \@url [1]{\endgroup\@href {#1}{\urlprefix }}%
\providecommand \urlprefix  [0]{URL }%
\providecommand \Eprint [0]{\href }%
\providecommand \doibase [0]{https://doi.org/}%
\providecommand \selectlanguage [0]{\@gobble}%
\providecommand \bibinfo  [0]{\@secondoftwo}%
\providecommand \bibfield  [0]{\@secondoftwo}%
\providecommand \translation [1]{[#1]}%
\providecommand \BibitemOpen [0]{}%
\providecommand \bibitemStop [0]{}%
\providecommand \bibitemNoStop [0]{.\EOS\space}%
\providecommand \EOS [0]{\spacefactor3000\relax}%
\providecommand \BibitemShut  [1]{\csname bibitem#1\endcsname}%
\let\auto@bib@innerbib\@empty
%</preamble>
\bibitem [{\citenamefont {Stohr}(2011)}]{stohr2011linac}%
  \BibitemOpen
  \bibfield  {author} {\bibinfo {author} {\bibfnamefont {J.}~\bibnamefont
  {Stohr}},\ }\href@noop {} {\enquote {\bibinfo {title} {Linac coherent light
  source ii {(LCLS-II)} conceptual design report},}\ }\bibinfo {type} {Tech.
  Rep.}\ (\bibinfo  {institution} {SLAC National Accelerator Lab., Menlo Park,
  CA (United States)},\ \bibinfo {year} {2011})\BibitemShut {NoStop}%
\bibitem [{\citenamefont {Galayda}()}]{Galayda_IPAC2014-TUOCA01}%
  \BibitemOpen
  \bibfield  {author} {\bibinfo {author} {\bibfnamefont {J.}~\bibnamefont
  {Galayda}},\ }\bibfield  {title} {\enquote {\bibinfo {title} {{T}he {L}inac
  {C}oherent {L}ight {S}ource{-II} {P}roject},}\ }in\ \href@noop {} {\emph
  {\bibinfo {booktitle} {Proc. 5th International Particle Accelerator
  Conference (IPAC'14), Dresden, Germany, June 15-20, 2014}}},\ \bibinfo
  {series and number} {\bibinfo {series} {International Particle Accelerator
  Conference}\ No.~\bibinfo {number} {5}}\ (\bibinfo  {publisher} {JACoW})\
  pp.\ \bibinfo {pages} {935--937}\BibitemShut {NoStop}%
\bibitem [{\citenamefont {Grassellino}\ \emph {et~al.}(2013)\citenamefont
  {Grassellino}, \citenamefont {Romanenko}, \citenamefont {Sergatskov},
  \citenamefont {Melnychuk}, \citenamefont {Trenikhina}, \citenamefont
  {Crawford}, \citenamefont {Rowe}, \citenamefont {Wong}, \citenamefont
  {Khabiboulline},\ and\ \citenamefont {Barkov}}]{grassellino2013nitrogen}%
  \BibitemOpen
  \bibfield  {author} {\bibinfo {author} {\bibfnamefont {A.}~\bibnamefont
  {Grassellino}}, \bibinfo {author} {\bibfnamefont {A.}~\bibnamefont
  {Romanenko}}, \bibinfo {author} {\bibfnamefont {D.}~\bibnamefont
  {Sergatskov}}, \bibinfo {author} {\bibfnamefont {O.}~\bibnamefont
  {Melnychuk}}, \bibinfo {author} {\bibfnamefont {Y.}~\bibnamefont
  {Trenikhina}}, \bibinfo {author} {\bibfnamefont {A.}~\bibnamefont
  {Crawford}}, \bibinfo {author} {\bibfnamefont {A.}~\bibnamefont {Rowe}},
  \bibinfo {author} {\bibfnamefont {M.}~\bibnamefont {Wong}}, \bibinfo {author}
  {\bibfnamefont {T.}~\bibnamefont {Khabiboulline}},\ and\ \bibinfo {author}
  {\bibfnamefont {F.}~\bibnamefont {Barkov}},\ }\bibfield  {title} {\enquote
  {\bibinfo {title} {Nitrogen and argon doping of niobium for superconducting
  radio frequency cavities: a pathway to highly efficient accelerating
  structures},}\ }\href@noop {} {\bibfield  {journal} {\bibinfo  {journal}
  {Superconductor Science and Technology}\ }\textbf {\bibinfo {volume} {26}},\
  \bibinfo {pages} {102001} (\bibinfo {year} {2013})}\BibitemShut {NoStop}%
\bibitem [{\citenamefont {Crawford}\ \emph {et~al.}(2014)\citenamefont
  {Crawford}, \citenamefont {Eichhorn}, \citenamefont {Furuta}, \citenamefont
  {Ge}, \citenamefont {Geng}, \citenamefont {Gonnella}, \citenamefont
  {Grassellino}, \citenamefont {Hocker}, \citenamefont {Hoffstaetter},
  \citenamefont {Liepe} \emph {et~al.}}]{crawford2014joint}%
  \BibitemOpen
  \bibfield  {author} {\bibinfo {author} {\bibfnamefont {A.}~\bibnamefont
  {Crawford}}, \bibinfo {author} {\bibfnamefont {R.}~\bibnamefont {Eichhorn}},
  \bibinfo {author} {\bibfnamefont {F.}~\bibnamefont {Furuta}}, \bibinfo
  {author} {\bibfnamefont {G.}~\bibnamefont {Ge}}, \bibinfo {author}
  {\bibfnamefont {R.}~\bibnamefont {Geng}}, \bibinfo {author} {\bibfnamefont
  {D.}~\bibnamefont {Gonnella}}, \bibinfo {author} {\bibfnamefont
  {A.}~\bibnamefont {Grassellino}}, \bibinfo {author} {\bibfnamefont
  {A.}~\bibnamefont {Hocker}}, \bibinfo {author} {\bibfnamefont
  {G.}~\bibnamefont {Hoffstaetter}}, \bibinfo {author} {\bibfnamefont
  {M.}~\bibnamefont {Liepe}}, \emph {et~al.},\ }\bibfield  {title} {\enquote
  {\bibinfo {title} {The joint high $\mathrm{Q_0}$ {R}\&{D} program for
  {LCLS-II}},}\ }\href@noop {} {\bibfield  {journal} {\bibinfo  {journal}
  {Proceedings of IPAC 2014}\ } (\bibinfo {year} {2014})}\BibitemShut {NoStop}%
\bibitem [{\citenamefont {Grassellino}\ \emph {et~al.}(2015)\citenamefont
  {Grassellino}, \citenamefont {Romanenko}, \citenamefont {Posen},
  \citenamefont {Trenikhina}, \citenamefont {Melnychuk}, \citenamefont
  {Sergatskov}, \citenamefont {Merio}, \citenamefont {Checchin},\ and\
  \citenamefont {Martinello}}]{grassellino2015n}%
  \BibitemOpen
  \bibfield  {author} {\bibinfo {author} {\bibfnamefont {A.}~\bibnamefont
  {Grassellino}}, \bibinfo {author} {\bibfnamefont {A.}~\bibnamefont
  {Romanenko}}, \bibinfo {author} {\bibfnamefont {S.}~\bibnamefont {Posen}},
  \bibinfo {author} {\bibfnamefont {Y.}~\bibnamefont {Trenikhina}}, \bibinfo
  {author} {\bibfnamefont {O.}~\bibnamefont {Melnychuk}}, \bibinfo {author}
  {\bibfnamefont {D.}~\bibnamefont {Sergatskov}}, \bibinfo {author}
  {\bibfnamefont {M.}~\bibnamefont {Merio}}, \bibinfo {author} {\bibfnamefont
  {M.}~\bibnamefont {Checchin}},\ and\ \bibinfo {author} {\bibfnamefont
  {M.}~\bibnamefont {Martinello}},\ }\bibfield  {title} {\enquote {\bibinfo
  {title} {N doping: progress in development and understanding},}\ }\href@noop
  {} {\bibfield  {journal} {\bibinfo  {journal} {Proc. SRF’2015}\ ,\ \bibinfo
  {pages} {48--54}} (\bibinfo {year} {2015})}\BibitemShut {NoStop}%
\bibitem [{\citenamefont {Gonnella}\ \emph {et~al.}(2018)\citenamefont
  {Gonnella}, \citenamefont {Aderhold}, \citenamefont {Burrill}, \citenamefont
  {Daly}, \citenamefont {Davis}, \citenamefont {Grassellino}, \citenamefont
  {Grimm}, \citenamefont {Khabiboulline}, \citenamefont {Marhauser},
  \citenamefont {Melnychuk} \emph {et~al.}}]{gonnella2018industrialization}%
  \BibitemOpen
  \bibfield  {author} {\bibinfo {author} {\bibfnamefont {D.}~\bibnamefont
  {Gonnella}}, \bibinfo {author} {\bibfnamefont {S.}~\bibnamefont {Aderhold}},
  \bibinfo {author} {\bibfnamefont {A.}~\bibnamefont {Burrill}}, \bibinfo
  {author} {\bibfnamefont {E.}~\bibnamefont {Daly}}, \bibinfo {author}
  {\bibfnamefont {K.}~\bibnamefont {Davis}}, \bibinfo {author} {\bibfnamefont
  {A.}~\bibnamefont {Grassellino}}, \bibinfo {author} {\bibfnamefont
  {C.}~\bibnamefont {Grimm}}, \bibinfo {author} {\bibfnamefont
  {T.}~\bibnamefont {Khabiboulline}}, \bibinfo {author} {\bibfnamefont
  {F.}~\bibnamefont {Marhauser}}, \bibinfo {author} {\bibfnamefont
  {O.}~\bibnamefont {Melnychuk}}, \emph {et~al.},\ }\bibfield  {title}
  {\enquote {\bibinfo {title} {Industrialization of the nitrogen-doping
  preparation for {SRF} cavities for {LCLS-II}},}\ }\href@noop {} {\bibfield
  {journal} {\bibinfo  {journal} {Nuclear Instruments and Methods in Physics
  Research Section A: Accelerators, Spectrometers, Detectors and Associated
  Equipment}\ }\textbf {\bibinfo {volume} {883}},\ \bibinfo {pages} {143--150}
  (\bibinfo {year} {2018})}\BibitemShut {NoStop}%
\bibitem [{\citenamefont {McNeil}\ and\ \citenamefont
  {Thompson}(2010)}]{mcneil2010x}%
  \BibitemOpen
  \bibfield  {author} {\bibinfo {author} {\bibfnamefont {B.~W.}\ \bibnamefont
  {McNeil}}\ and\ \bibinfo {author} {\bibfnamefont {N.~R.}\ \bibnamefont
  {Thompson}},\ }\bibfield  {title} {\enquote {\bibinfo {title} {X-ray
  free-electron lasers},}\ }\href@noop {} {\bibfield  {journal} {\bibinfo
  {journal} {Nature photonics}\ }\textbf {\bibinfo {volume} {4}},\ \bibinfo
  {pages} {814--821} (\bibinfo {year} {2010})}\BibitemShut {NoStop}%
\bibitem [{\citenamefont {Kao}(2020)}]{kao2020challenges}%
  \BibitemOpen
  \bibfield  {author} {\bibinfo {author} {\bibfnamefont {C.-C.}\ \bibnamefont
  {Kao}},\ }\bibfield  {title} {\enquote {\bibinfo {title} {Challenges and
  opportunities for the next decade of {XFELs}},}\ }\href@noop {} {\bibfield
  {journal} {\bibinfo  {journal} {Nature Reviews Physics}\ ,\ \bibinfo {pages}
  {1--2}} (\bibinfo {year} {2020})}\BibitemShut {NoStop}%
\bibitem [{\citenamefont {Cornacchia}(1998)}]{cornacchia1998linac}%
  \BibitemOpen
  \bibfield  {author} {\bibinfo {author} {\bibfnamefont {M.}~\bibnamefont
  {Cornacchia}},\ }\href@noop {} {\enquote {\bibinfo {title} {Linac coherent
  light source {(LCLS)} design study report},}\ }\bibinfo {type} {Tech. Rep.}\
  (\bibinfo  {institution} {Stanford Linear Accelerator Center, Menlo Park, CA
  (US)},\ \bibinfo {year} {1998})\BibitemShut {NoStop}%
\bibitem [{\citenamefont {Schoenlein}\ \emph {et~al.}(2017)\citenamefont
  {Schoenlein}, \citenamefont {Boutet}, \citenamefont {Minitti},\ and\
  \citenamefont {Dunne}}]{schoenlein2017linac}%
  \BibitemOpen
  \bibfield  {author} {\bibinfo {author} {\bibfnamefont {R.}~\bibnamefont
  {Schoenlein}}, \bibinfo {author} {\bibfnamefont {S.}~\bibnamefont {Boutet}},
  \bibinfo {author} {\bibfnamefont {M.}~\bibnamefont {Minitti}},\ and\ \bibinfo
  {author} {\bibfnamefont {A.}~\bibnamefont {Dunne}},\ }\bibfield  {title}
  {\enquote {\bibinfo {title} {The linac coherent light source: recent
  developments and future plans},}\ }\href@noop {} {\bibfield  {journal}
  {\bibinfo  {journal} {Applied Sciences}\ }\textbf {\bibinfo {volume} {7}},\
  \bibinfo {pages} {850} (\bibinfo {year} {2017})}\BibitemShut {NoStop}%
\bibitem [{\citenamefont {Padamsee}, \citenamefont {Knobloch},\ and\
  \citenamefont {Hays}(1998)}]{padamsee1998rf}%
  \BibitemOpen
  \bibfield  {author} {\bibinfo {author} {\bibfnamefont {H.}~\bibnamefont
  {Padamsee}}, \bibinfo {author} {\bibfnamefont {J.}~\bibnamefont {Knobloch}},\
  and\ \bibinfo {author} {\bibfnamefont {T.}~\bibnamefont {Hays}},\ }\href@noop
  {} {\emph {\bibinfo {title} {RF superconductivity for accelerators}}}\
  (\bibinfo  {publisher} {John Wiley \& Sons, New York},\ \bibinfo {year}
  {1998})\BibitemShut {NoStop}%
\bibitem [{\citenamefont {Fowler}\ and\ \citenamefont
  {Nordheim}(1928)}]{fowler1928electron}%
  \BibitemOpen
  \bibfield  {author} {\bibinfo {author} {\bibfnamefont {R.~H.}\ \bibnamefont
  {Fowler}}\ and\ \bibinfo {author} {\bibfnamefont {L.}~\bibnamefont
  {Nordheim}},\ }\bibfield  {title} {\enquote {\bibinfo {title} {Electron
  emission in intense electric fields},}\ }\href@noop {} {\bibfield  {journal}
  {\bibinfo  {journal} {Proceedings of the Royal Society of London. Series A,
  Containing Papers of a Mathematical and Physical Character}\ }\textbf
  {\bibinfo {volume} {119}},\ \bibinfo {pages} {173--181} (\bibinfo {year}
  {1928})}\BibitemShut {NoStop}%
\bibitem [{\citenamefont {Bagus}\ \emph {et~al.}(2008)\citenamefont {Bagus},
  \citenamefont {K{\"a}fer}, \citenamefont {Witte},\ and\ \citenamefont
  {W{\"o}ll}}]{bagus2008work}%
  \BibitemOpen
  \bibfield  {author} {\bibinfo {author} {\bibfnamefont {P.~S.}\ \bibnamefont
  {Bagus}}, \bibinfo {author} {\bibfnamefont {D.}~\bibnamefont {K{\"a}fer}},
  \bibinfo {author} {\bibfnamefont {G.}~\bibnamefont {Witte}},\ and\ \bibinfo
  {author} {\bibfnamefont {C.}~\bibnamefont {W{\"o}ll}},\ }\bibfield  {title}
  {\enquote {\bibinfo {title} {Work function changes induced by charged
  adsorbates: origin of the polarity asymmetry},}\ }\href@noop {} {\bibfield
  {journal} {\bibinfo  {journal} {Physical review letters}\ }\textbf {\bibinfo
  {volume} {100}},\ \bibinfo {pages} {126101} (\bibinfo {year}
  {2008})}\BibitemShut {NoStop}%
\bibitem [{\citenamefont {Doleans}\ \emph
  {et~al.}(2016{\natexlab{a}})\citenamefont {Doleans}, \citenamefont {Tyagi},
  \citenamefont {Afanador}, \citenamefont {McMahan}, \citenamefont {Ball},
  \citenamefont {Barnhart}, \citenamefont {Blokland}, \citenamefont {Crofford},
  \citenamefont {Degraff}, \citenamefont {Gold} \emph
  {et~al.}}]{doleans2016situ}%
  \BibitemOpen
  \bibfield  {author} {\bibinfo {author} {\bibfnamefont {M.}~\bibnamefont
  {Doleans}}, \bibinfo {author} {\bibfnamefont {P.}~\bibnamefont {Tyagi}},
  \bibinfo {author} {\bibfnamefont {R.}~\bibnamefont {Afanador}}, \bibinfo
  {author} {\bibfnamefont {C.}~\bibnamefont {McMahan}}, \bibinfo {author}
  {\bibfnamefont {J.}~\bibnamefont {Ball}}, \bibinfo {author} {\bibfnamefont
  {D.}~\bibnamefont {Barnhart}}, \bibinfo {author} {\bibfnamefont
  {W.}~\bibnamefont {Blokland}}, \bibinfo {author} {\bibfnamefont
  {M.}~\bibnamefont {Crofford}}, \bibinfo {author} {\bibfnamefont
  {B.}~\bibnamefont {Degraff}}, \bibinfo {author} {\bibfnamefont
  {S.}~\bibnamefont {Gold}}, \emph {et~al.},\ }\bibfield  {title} {\enquote
  {\bibinfo {title} {In-situ plasma processing to increase the accelerating
  gradients of superconducting radio-frequency cavities},}\ }\href@noop {}
  {\bibfield  {journal} {\bibinfo  {journal} {Nuclear Instruments and Methods
  in Physics Research Section A: Accelerators, Spectrometers, Detectors and
  Associated Equipment}\ }\textbf {\bibinfo {volume} {812}},\ \bibinfo {pages}
  {50--59} (\bibinfo {year} {2016}{\natexlab{a}})}\BibitemShut {NoStop}%
\bibitem [{\citenamefont {Cao}\ \emph {et~al.}(2013)\citenamefont {Cao},
  \citenamefont {Ford}, \citenamefont {Bishnoi}, \citenamefont {Proslier},
  \citenamefont {Albee}, \citenamefont {Hommerding}, \citenamefont
  {Korczakowski}, \citenamefont {Cooley}, \citenamefont {Ciovati},\ and\
  \citenamefont {Zasadzinski}}]{cao2013}%
  \BibitemOpen
  \bibfield  {author} {\bibinfo {author} {\bibfnamefont {C.}~\bibnamefont
  {Cao}}, \bibinfo {author} {\bibfnamefont {D.}~\bibnamefont {Ford}}, \bibinfo
  {author} {\bibfnamefont {S.}~\bibnamefont {Bishnoi}}, \bibinfo {author}
  {\bibfnamefont {T.}~\bibnamefont {Proslier}}, \bibinfo {author}
  {\bibfnamefont {B.}~\bibnamefont {Albee}}, \bibinfo {author} {\bibfnamefont
  {E.}~\bibnamefont {Hommerding}}, \bibinfo {author} {\bibfnamefont
  {A.}~\bibnamefont {Korczakowski}}, \bibinfo {author} {\bibfnamefont
  {L.}~\bibnamefont {Cooley}}, \bibinfo {author} {\bibfnamefont
  {G.}~\bibnamefont {Ciovati}},\ and\ \bibinfo {author} {\bibfnamefont {J.~F.}\
  \bibnamefont {Zasadzinski}},\ }\bibfield  {title} {\enquote {\bibinfo {title}
  {Detection of surface carbon and hydrocarbons in hot spot regions of niobium
  {SRF} cavities by {Raman} spectroscopy},}\ }\href@noop {} {\bibfield
  {journal} {\bibinfo  {journal} {Physical Review Special Topics - Accelerators
  and Beams}\ }\textbf {\bibinfo {volume} {25}} (\bibinfo {year}
  {2013})}\BibitemShut {NoStop}%
\bibitem [{\citenamefont {Pudasaini}\ \emph {et~al.}(2020)\citenamefont
  {Pudasaini}, \citenamefont {Eremeev}, \citenamefont {Reece}, \citenamefont
  {Tuggle},\ and\ \citenamefont {Kelley}}]{pudasaini2020analysis}%
  \BibitemOpen
  \bibfield  {author} {\bibinfo {author} {\bibfnamefont {U.}~\bibnamefont
  {Pudasaini}}, \bibinfo {author} {\bibfnamefont {G.~V.}\ \bibnamefont
  {Eremeev}}, \bibinfo {author} {\bibfnamefont {C.~E.}\ \bibnamefont {Reece}},
  \bibinfo {author} {\bibfnamefont {J.}~\bibnamefont {Tuggle}},\ and\ \bibinfo
  {author} {\bibfnamefont {M.~J.}\ \bibnamefont {Kelley}},\ }\bibfield  {title}
  {\enquote {\bibinfo {title} {{Analysis of {RF} losses and material
  characterization of samples removed from a \ch{Nb_3Sn}-coated superconducting
  {RF} cavity}},}\ }\href@noop {} {\bibfield  {journal} {\bibinfo  {journal}
  {Superconductor Science and Technology}\ }\textbf {\bibinfo {volume} {33}},\
  \bibinfo {pages} {045012} (\bibinfo {year} {2020})}\BibitemShut {NoStop}%
\bibitem [{\citenamefont {Mason}\ \emph {et~al.}(2006)\citenamefont {Mason},
  \citenamefont {Abernathy}, \citenamefont {Anderson}, \citenamefont {Ankner},
  \citenamefont {Egami}, \citenamefont {Ehlers}, \citenamefont {Ekkebus},
  \citenamefont {Granroth}, \citenamefont {Hagen}, \citenamefont {Herwig} \emph
  {et~al.}}]{mason2006spallation}%
  \BibitemOpen
  \bibfield  {author} {\bibinfo {author} {\bibfnamefont {T.}~\bibnamefont
  {Mason}}, \bibinfo {author} {\bibfnamefont {D.}~\bibnamefont {Abernathy}},
  \bibinfo {author} {\bibfnamefont {I.}~\bibnamefont {Anderson}}, \bibinfo
  {author} {\bibfnamefont {J.}~\bibnamefont {Ankner}}, \bibinfo {author}
  {\bibfnamefont {T.}~\bibnamefont {Egami}}, \bibinfo {author} {\bibfnamefont
  {G.}~\bibnamefont {Ehlers}}, \bibinfo {author} {\bibfnamefont
  {A.}~\bibnamefont {Ekkebus}}, \bibinfo {author} {\bibfnamefont
  {G.}~\bibnamefont {Granroth}}, \bibinfo {author} {\bibfnamefont
  {M.}~\bibnamefont {Hagen}}, \bibinfo {author} {\bibfnamefont
  {K.}~\bibnamefont {Herwig}}, \emph {et~al.},\ }\bibfield  {title} {\enquote
  {\bibinfo {title} {{The Spallation Neutron Source in Oak Ridge: A powerful
  tool for materials research}},}\ }\href@noop {} {\bibfield  {journal}
  {\bibinfo  {journal} {Physica B: Condensed Matter}\ }\textbf {\bibinfo
  {volume} {385}},\ \bibinfo {pages} {955--960} (\bibinfo {year}
  {2006})}\BibitemShut {NoStop}%
\bibitem [{\citenamefont {Doleans}(2016)}]{doleans2016ignition}%
  \BibitemOpen
  \bibfield  {author} {\bibinfo {author} {\bibfnamefont {M.}~\bibnamefont
  {Doleans}},\ }\bibfield  {title} {\enquote {\bibinfo {title} {Ignition and
  monitoring technique for plasma processing of multicell superconducting
  radio-frequency cavities},}\ }\href@noop {} {\bibfield  {journal} {\bibinfo
  {journal} {Journal of Applied Physics}\ }\textbf {\bibinfo {volume} {120}},\
  \bibinfo {pages} {243301} (\bibinfo {year} {2016})}\BibitemShut {NoStop}%
\bibitem [{\citenamefont {Doleans}\ \emph
  {et~al.}(2016{\natexlab{b}})\citenamefont {Doleans}, \citenamefont
  {Vandygriff}, \citenamefont {Stewart}, \citenamefont {Gold}, \citenamefont
  {Neustadt}, \citenamefont {Strong}, \citenamefont {McMahan}, \citenamefont
  {Lee}, \citenamefont {Tyagi}, \citenamefont {Vandygriff} \emph
  {et~al.}}]{doleans2016plasma}%
  \BibitemOpen
  \bibfield  {author} {\bibinfo {author} {\bibfnamefont {M.}~\bibnamefont
  {Doleans}}, \bibinfo {author} {\bibfnamefont {D.}~\bibnamefont {Vandygriff}},
  \bibinfo {author} {\bibfnamefont {S.}~\bibnamefont {Stewart}}, \bibinfo
  {author} {\bibfnamefont {S.}~\bibnamefont {Gold}}, \bibinfo {author}
  {\bibfnamefont {T.}~\bibnamefont {Neustadt}}, \bibinfo {author}
  {\bibfnamefont {W.}~\bibnamefont {Strong}}, \bibinfo {author} {\bibfnamefont
  {C.}~\bibnamefont {McMahan}}, \bibinfo {author} {\bibfnamefont {S.-W.}\
  \bibnamefont {Lee}}, \bibinfo {author} {\bibfnamefont {P.}~\bibnamefont
  {Tyagi}}, \bibinfo {author} {\bibfnamefont {D.}~\bibnamefont {Vandygriff}},
  \emph {et~al.},\ }\bibfield  {title} {\enquote {\bibinfo {title} {Plasma
  processing to improve the performance of the {SNS} superconducting linac},}\
  }\href@noop {} {\bibfield  {journal} {\bibinfo  {journal} {Proceedings, 28th
  International Conference on RF Superconductivity (LINAC16), East Lansing, MI,
  USA}\ } (\bibinfo {year} {2016}{\natexlab{b}})},\ \bibinfo {note}
  {{WE2A03}}\BibitemShut {NoStop}%
\bibitem [{\citenamefont {Aune}\ \emph {et~al.}(2000)\citenamefont {Aune},
  \citenamefont {Bandelmann}, \citenamefont {Bloess}, \citenamefont {Bonin},
  \citenamefont {Bosotti}, \citenamefont {Champion}, \citenamefont {Crawford},
  \citenamefont {Deppe}, \citenamefont {Dwersteg}, \citenamefont {Edwards}
  \emph {et~al.}}]{aune2000superconducting}%
  \BibitemOpen
  \bibfield  {author} {\bibinfo {author} {\bibfnamefont {B.}~\bibnamefont
  {Aune}}, \bibinfo {author} {\bibfnamefont {R.}~\bibnamefont {Bandelmann}},
  \bibinfo {author} {\bibfnamefont {D.}~\bibnamefont {Bloess}}, \bibinfo
  {author} {\bibfnamefont {B.}~\bibnamefont {Bonin}}, \bibinfo {author}
  {\bibfnamefont {A.}~\bibnamefont {Bosotti}}, \bibinfo {author} {\bibfnamefont
  {M.}~\bibnamefont {Champion}}, \bibinfo {author} {\bibfnamefont
  {C.}~\bibnamefont {Crawford}}, \bibinfo {author} {\bibfnamefont
  {G.}~\bibnamefont {Deppe}}, \bibinfo {author} {\bibfnamefont
  {B.}~\bibnamefont {Dwersteg}}, \bibinfo {author} {\bibfnamefont
  {D.}~\bibnamefont {Edwards}}, \emph {et~al.},\ }\bibfield  {title} {\enquote
  {\bibinfo {title} {Superconducting {TESLA} cavities},}\ }\href@noop {}
  {\bibfield  {journal} {\bibinfo  {journal} {Physical Review Special
  Topics-Accelerators and Beams}\ }\textbf {\bibinfo {volume} {3}},\ \bibinfo
  {pages} {092001} (\bibinfo {year} {2000})}\BibitemShut {NoStop}%
\bibitem [{\citenamefont {Berrutti}\ \emph {et~al.}(2018)\citenamefont
  {Berrutti}, \citenamefont {Doleans}, \citenamefont {Gonnella}, \citenamefont
  {Grassellino}, \citenamefont {Khabiboulline}, \citenamefont {Kim},
  \citenamefont {Lanza}, \citenamefont {Martinello}, \citenamefont {Ross},\
  and\ \citenamefont {Tippey}}]{berrutti2018}%
  \BibitemOpen
  \bibfield  {author} {\bibinfo {author} {\bibfnamefont {P.}~\bibnamefont
  {Berrutti}}, \bibinfo {author} {\bibfnamefont {M.}~\bibnamefont {Doleans}},
  \bibinfo {author} {\bibfnamefont {D.}~\bibnamefont {Gonnella}}, \bibinfo
  {author} {\bibfnamefont {A.}~\bibnamefont {Grassellino}}, \bibinfo {author}
  {\bibfnamefont {T.}~\bibnamefont {Khabiboulline}}, \bibinfo {author}
  {\bibfnamefont {S.}~\bibnamefont {Kim}}, \bibinfo {author} {\bibfnamefont
  {G.}~\bibnamefont {Lanza}}, \bibinfo {author} {\bibfnamefont
  {M.}~\bibnamefont {Martinello}}, \bibinfo {author} {\bibfnamefont
  {M.}~\bibnamefont {Ross}},\ and\ \bibinfo {author} {\bibfnamefont
  {K.}~\bibnamefont {Tippey}},\ }\bibfield  {title} {\enquote {\bibinfo {title}
  {{U}pdate on {P}lasma {P}rocessing {R}\&{D} for {LCLS-II}},}\ }in\ \href@noop
  {} {\emph {\bibinfo {booktitle} {Proc. 9th International Particle Accelerator
  Conference (IPAC'18), Vancouver, BC, Canada}}}\ (\bibinfo {address} {Geneva,
  Switzerland},\ \bibinfo {year} {2018})\BibitemShut {NoStop}%
\bibitem [{\citenamefont {Berrutti}\ \emph {et~al.}(2019)\citenamefont
  {Berrutti}, \citenamefont {Giaccone}, \citenamefont {Martinello},
  \citenamefont {Grassellino}, \citenamefont {Khabiboulline}, \citenamefont
  {Doleans}, \citenamefont {Kim}, \citenamefont {Gonnella}, \citenamefont
  {Lanza},\ and\ \citenamefont {Ross}}]{berrutti2019plasma}%
  \BibitemOpen
  \bibfield  {author} {\bibinfo {author} {\bibfnamefont {P.}~\bibnamefont
  {Berrutti}}, \bibinfo {author} {\bibfnamefont {B.}~\bibnamefont {Giaccone}},
  \bibinfo {author} {\bibfnamefont {M.}~\bibnamefont {Martinello}}, \bibinfo
  {author} {\bibfnamefont {A.}~\bibnamefont {Grassellino}}, \bibinfo {author}
  {\bibfnamefont {T.}~\bibnamefont {Khabiboulline}}, \bibinfo {author}
  {\bibfnamefont {M.}~\bibnamefont {Doleans}}, \bibinfo {author} {\bibfnamefont
  {S.}~\bibnamefont {Kim}}, \bibinfo {author} {\bibfnamefont {D.}~\bibnamefont
  {Gonnella}}, \bibinfo {author} {\bibfnamefont {G.}~\bibnamefont {Lanza}},\
  and\ \bibinfo {author} {\bibfnamefont {M.}~\bibnamefont {Ross}},\ }\bibfield
  {title} {\enquote {\bibinfo {title} {Plasma ignition and detection for
  in-situ cleaning of $1.3$ {GHz} 9-cell cavities},}\ }\href@noop {} {\bibfield
   {journal} {\bibinfo  {journal} {Journal of Applied Physics}\ }\textbf
  {\bibinfo {volume} {126}},\ \bibinfo {pages} {023302} (\bibinfo {year}
  {2019})}\BibitemShut {NoStop}%
\bibitem [{\citenamefont {Giaccone}\ \emph {et~al.}(2019)\citenamefont
  {Giaccone}, \citenamefont {Berrutti}, \citenamefont {Doleans}, \citenamefont
  {Gonnella}, \citenamefont {Grassellino}, \citenamefont {Lanza}, \citenamefont
  {Martinello}, \citenamefont {Ross},\ and\ \citenamefont
  {Zasadzinski}}]{giaccone_srf2019-frcab7}%
  \BibitemOpen
  \bibfield  {author} {\bibinfo {author} {\bibfnamefont {B.}~\bibnamefont
  {Giaccone}}, \bibinfo {author} {\bibfnamefont {P.}~\bibnamefont {Berrutti}},
  \bibinfo {author} {\bibfnamefont {M.}~\bibnamefont {Doleans}}, \bibinfo
  {author} {\bibfnamefont {D.}~\bibnamefont {Gonnella}}, \bibinfo {author}
  {\bibfnamefont {A.}~\bibnamefont {Grassellino}}, \bibinfo {author}
  {\bibfnamefont {G.}~\bibnamefont {Lanza}}, \bibinfo {author} {\bibfnamefont
  {M.}~\bibnamefont {Martinello}}, \bibinfo {author} {\bibfnamefont
  {M.}~\bibnamefont {Ross}},\ and\ \bibinfo {author} {\bibfnamefont
  {J.}~\bibnamefont {Zasadzinski}},\ }\bibfield  {title} {\enquote {\bibinfo
  {title} {Plasma processing to reduce field emission in {LCLS-II} 1.3 {GHz}
  {SRF} cavities},}\ }in\ \href@noop {} {\emph {\bibinfo {booktitle} {Proc.
  SRF'19}}},\ \bibinfo {series and number} {\bibinfo {series} {International
  Conference on RF Superconductivity}\ No.~\bibinfo {number} {19}}\ (\bibinfo
  {publisher} {JACoW Publishing, Geneva, Switzerland},\ \bibinfo {year}
  {2019})\ pp.\ \bibinfo {pages} {1231--1238}\BibitemShut {NoStop}%
\bibitem [{\citenamefont {Wu}\ \emph {et~al.}(2018)\citenamefont {Wu},
  \citenamefont {Yang}, \citenamefont {Hu}, \citenamefont {Li}, \citenamefont
  {Huang}, \citenamefont {Li}, \citenamefont {Chu}, \citenamefont {Xiong},
  \citenamefont {Guo}, \citenamefont {Yue},\ and\ \citenamefont
  {Hu}}]{wu2018situ}%
  \BibitemOpen
  \bibfield  {author} {\bibinfo {author} {\bibfnamefont {A.}~\bibnamefont
  {Wu}}, \bibinfo {author} {\bibfnamefont {L.}~\bibnamefont {Yang}}, \bibinfo
  {author} {\bibfnamefont {C.}~\bibnamefont {Hu}}, \bibinfo {author}
  {\bibfnamefont {C.}~\bibnamefont {Li}}, \bibinfo {author} {\bibfnamefont
  {S.}~\bibnamefont {Huang}}, \bibinfo {author} {\bibfnamefont
  {Y.}~\bibnamefont {Li}}, \bibinfo {author} {\bibfnamefont {Q.}~\bibnamefont
  {Chu}}, \bibinfo {author} {\bibfnamefont {P.}~\bibnamefont {Xiong}}, \bibinfo
  {author} {\bibfnamefont {H.}~\bibnamefont {Guo}}, \bibinfo {author}
  {\bibfnamefont {W.}~\bibnamefont {Yue}},\ and\ \bibinfo {author}
  {\bibfnamefont {Y.}~\bibnamefont {Hu}},\ }\bibfield  {title} {\enquote
  {\bibinfo {title} {In-situ plasma cleaning to decrease the field emission
  effect of half-wave superconducting radio-frequency cavities},}\ }\href@noop
  {} {\bibfield  {journal} {\bibinfo  {journal} {Nuclear Instruments and
  Methods in Physics Research Section A: Accelerators, Spectrometers, Detectors
  and Associated Equipment}\ }\textbf {\bibinfo {volume} {905}},\ \bibinfo
  {pages} {61--70} (\bibinfo {year} {2018})}\BibitemShut {NoStop}%
\bibitem [{\citenamefont {Wu}\ \emph {et~al.}(2019)\citenamefont {Wu},
  \citenamefont {Huang}, \citenamefont {Li}, \citenamefont {Chu}, \citenamefont
  {Guo}, \citenamefont {Xiong}, \citenamefont {Song}, \citenamefont {Pan},
  \citenamefont {Tan}, \citenamefont {Yue},\ and\ \citenamefont
  {Zhang}}]{wu2019cryostat}%
  \BibitemOpen
  \bibfield  {author} {\bibinfo {author} {\bibfnamefont {A.}~\bibnamefont
  {Wu}}, \bibinfo {author} {\bibfnamefont {S.}~\bibnamefont {Huang}}, \bibinfo
  {author} {\bibfnamefont {C.}~\bibnamefont {Li}}, \bibinfo {author}
  {\bibfnamefont {Q.}~\bibnamefont {Chu}}, \bibinfo {author} {\bibfnamefont
  {H.}~\bibnamefont {Guo}}, \bibinfo {author} {\bibfnamefont {P.}~\bibnamefont
  {Xiong}}, \bibinfo {author} {\bibfnamefont {Y.}~\bibnamefont {Song}},
  \bibinfo {author} {\bibfnamefont {F.}~\bibnamefont {Pan}}, \bibinfo {author}
  {\bibfnamefont {T.}~\bibnamefont {Tan}}, \bibinfo {author} {\bibfnamefont
  {W.}~\bibnamefont {Yue}},\ and\ \bibinfo {author} {\bibfnamefont
  {S.}~\bibnamefont {Zhang}},\ }\bibfield  {title} {\enquote {\bibinfo {title}
  {The cryostat results of carbon contamination and plasma cleaning for the
  field emission on the {SRF} cavity},}\ }in\ \href@noop {} {\emph {\bibinfo
  {booktitle} {19th Int. Conf. on RF Superconductivity (SRF'19), Dresden,
  Germany, 30 June-05 July 2019}}}\ (\bibinfo {organization} {JACOW Publishing,
  Geneva, Switzerland},\ \bibinfo {year} {2019})\ pp.\ \bibinfo {pages}
  {1038--1040}\BibitemShut {NoStop}%
\bibitem [{\citenamefont {Huang}\ \emph {et~al.}(2019)\citenamefont {Huang},
  \citenamefont {Chu}, \citenamefont {He}, \citenamefont {Li}, \citenamefont
  {Wu},\ and\ \citenamefont {Zhang}}]{huang2019effect}%
  \BibitemOpen
  \bibfield  {author} {\bibinfo {author} {\bibfnamefont {S.}~\bibnamefont
  {Huang}}, \bibinfo {author} {\bibfnamefont {Q.}~\bibnamefont {Chu}}, \bibinfo
  {author} {\bibfnamefont {Y.}~\bibnamefont {He}}, \bibinfo {author}
  {\bibfnamefont {C.}~\bibnamefont {Li}}, \bibinfo {author} {\bibfnamefont
  {A.}~\bibnamefont {Wu}},\ and\ \bibinfo {author} {\bibfnamefont
  {S.}~\bibnamefont {Zhang}},\ }\bibfield  {title} {\enquote {\bibinfo {title}
  {The effect of helium processing and plasma cleaning for low beta {HWR}
  cavity},}\ }in\ \href@noop {} {\emph {\bibinfo {booktitle} {19th Int. Conf.
  on RF Superconductivity (SRF'19), Dresden, Germany, 30 June-05 July 2019}}}\
  (\bibinfo {organization} {JACOW Publishing, Geneva, Switzerland},\ \bibinfo
  {year} {2019})\ pp.\ \bibinfo {pages} {1228--1230}\BibitemShut {NoStop}%
\bibitem [{\citenamefont {Brown}\ \emph {et~al.}(1966)\citenamefont {Brown}
  \emph {et~al.}}]{brown1966introduction}%
  \BibitemOpen
  \bibfield  {author} {\bibinfo {author} {\bibfnamefont {S.~C.}\ \bibnamefont
  {Brown}} \emph {et~al.},\ }\href@noop {} {\emph {\bibinfo {title}
  {Introduction to electrical discharges in gases}}}\ (\bibinfo  {publisher}
  {John Wiley \& Sons, New York},\ \bibinfo {year} {1966})\BibitemShut
  {NoStop}%
\bibitem [{\citenamefont {Fitzpatrick}(2014)}]{fitzpatrick2014plasma}%
  \BibitemOpen
  \bibfield  {author} {\bibinfo {author} {\bibfnamefont {R.}~\bibnamefont
  {Fitzpatrick}},\ }\href@noop {} {\emph {\bibinfo {title} {Plasma physics: an
  introduction}}}\ (\bibinfo  {publisher} {Crc Press},\ \bibinfo {year}
  {2014})\BibitemShut {NoStop}%
\bibitem [{\citenamefont {Tyagi}\ \emph {et~al.}(2014)\citenamefont {Tyagi},
  \citenamefont {Afanador}, \citenamefont {Doleans}, \citenamefont {McMahan},\
  and\ \citenamefont {Kim}}]{tyagi2014plasma}%
  \BibitemOpen
  \bibfield  {author} {\bibinfo {author} {\bibfnamefont {P.}~\bibnamefont
  {Tyagi}}, \bibinfo {author} {\bibfnamefont {R.}~\bibnamefont {Afanador}},
  \bibinfo {author} {\bibfnamefont {M.}~\bibnamefont {Doleans}}, \bibinfo
  {author} {\bibfnamefont {C.}~\bibnamefont {McMahan}},\ and\ \bibinfo {author}
  {\bibfnamefont {S.-H.}\ \bibnamefont {Kim}},\ }\bibfield  {title} {\enquote
  {\bibinfo {title} {Plasma processing of \ch{Nb} surfaces for {SRF}
  cavities},}\ }\href@noop {} {\bibfield  {journal} {\bibinfo  {journal}
  {Proceedings, 27th International Conference on RF Superconductivity
  (LINAC14), Geneva, Switzerland}\ } (\bibinfo {year} {2014})},\ \bibinfo
  {note} {{MOPP115}}\BibitemShut {NoStop}%
\bibitem [{lcl()}]{lclsIIdesign}%
  \BibitemOpen
  \href@noop {} {\enquote {\bibinfo {title} {{LCLS-II Final Design Report,
  LCLSII-1.1-DR-0251-R0}},}\ }\BibitemShut {NoStop}%
\bibitem [{\citenamefont {Doleans}\ \emph {et~al.}(2013)\citenamefont
  {Doleans}, \citenamefont {Afanador}, \citenamefont {Ball}, \citenamefont
  {Blokland}, \citenamefont {Crofford}, \citenamefont {Degraff}, \citenamefont
  {Douglas}, \citenamefont {Hannah}, \citenamefont {Howell}, \citenamefont
  {Kim} \emph {et~al.}}]{doleans2013plasma}%
  \BibitemOpen
  \bibfield  {author} {\bibinfo {author} {\bibfnamefont {M.}~\bibnamefont
  {Doleans}}, \bibinfo {author} {\bibfnamefont {R.}~\bibnamefont {Afanador}},
  \bibinfo {author} {\bibfnamefont {J.}~\bibnamefont {Ball}}, \bibinfo {author}
  {\bibfnamefont {W.}~\bibnamefont {Blokland}}, \bibinfo {author}
  {\bibfnamefont {M.}~\bibnamefont {Crofford}}, \bibinfo {author}
  {\bibfnamefont {B.}~\bibnamefont {Degraff}}, \bibinfo {author} {\bibfnamefont
  {D.}~\bibnamefont {Douglas}}, \bibinfo {author} {\bibfnamefont
  {B.}~\bibnamefont {Hannah}}, \bibinfo {author} {\bibfnamefont
  {M.}~\bibnamefont {Howell}}, \bibinfo {author} {\bibfnamefont
  {S.}~\bibnamefont {Kim}}, \emph {et~al.},\ }\bibfield  {title} {\enquote
  {\bibinfo {title} {{Plasma processing {R}\&{D} for the SNS superconducting
  linac {RF} cavities}},}\ \ }(\bibinfo {year} {2013})\BibitemShut {NoStop}%
\bibitem [{\citenamefont {Slater}(1946)}]{slater1946microwave}%
  \BibitemOpen
  \bibfield  {author} {\bibinfo {author} {\bibfnamefont {J.}~\bibnamefont
  {Slater}},\ }\bibfield  {title} {\enquote {\bibinfo {title} {Microwave
  electronics},}\ }\href@noop {} {\bibfield  {journal} {\bibinfo  {journal}
  {Reviews of Modern Physics}\ }\textbf {\bibinfo {volume} {18}},\ \bibinfo
  {pages} {441} (\bibinfo {year} {1946})}\BibitemShut {NoStop}%
\bibitem [{\citenamefont {Melnychuk}, \citenamefont {Grassellino},\ and\
  \citenamefont {Romanenko}(2014)}]{melnychuk2014error}%
  \BibitemOpen
  \bibfield  {author} {\bibinfo {author} {\bibfnamefont {O.}~\bibnamefont
  {Melnychuk}}, \bibinfo {author} {\bibfnamefont {A.}~\bibnamefont
  {Grassellino}},\ and\ \bibinfo {author} {\bibfnamefont {A.}~\bibnamefont
  {Romanenko}},\ }\bibfield  {title} {\enquote {\bibinfo {title} {Error
  analysis for intrinsic quality factor measurement in superconducting radio
  frequency resonators},}\ }\href@noop {} {\bibfield  {journal} {\bibinfo
  {journal} {Review of Scientific Instruments}\ }\textbf {\bibinfo {volume}
  {85}},\ \bibinfo {pages} {124705} (\bibinfo {year} {2014})}\BibitemShut
  {NoStop}%
\bibitem [{\citenamefont {Krueger}\ and\ \citenamefont
  {Larson}(2002)}]{krueger2002chipmunk}%
  \BibitemOpen
  \bibfield  {author} {\bibinfo {author} {\bibfnamefont {F.}~\bibnamefont
  {Krueger}}\ and\ \bibinfo {author} {\bibfnamefont {J.}~\bibnamefont
  {Larson}},\ }\bibfield  {title} {\enquote {\bibinfo {title} {Chipmunk {IV}:
  development of and experience with a new generation of radiation area
  monitors for accelerator applications},}\ }\href@noop {} {\bibfield
  {journal} {\bibinfo  {journal} {Nuclear Instruments and Methods in Physics
  Research Section A: Accelerators, Spectrometers, Detectors and Associated
  Equipment}\ }\textbf {\bibinfo {volume} {495}},\ \bibinfo {pages} {20--28}
  (\bibinfo {year} {2002})}\BibitemShut {NoStop}%
\bibitem [{\citenamefont {Martinello}\ \emph {et~al.}(2016)\citenamefont
  {Martinello}, \citenamefont {Grassellino}, \citenamefont {Checchin},
  \citenamefont {Romanenko}, \citenamefont {Melnychuk}, \citenamefont
  {Sergatskov}, \citenamefont {Posen},\ and\ \citenamefont
  {Zasadzinski}}]{martinello2016effect}%
  \BibitemOpen
  \bibfield  {author} {\bibinfo {author} {\bibfnamefont {M.}~\bibnamefont
  {Martinello}}, \bibinfo {author} {\bibfnamefont {A.}~\bibnamefont
  {Grassellino}}, \bibinfo {author} {\bibfnamefont {M.}~\bibnamefont
  {Checchin}}, \bibinfo {author} {\bibfnamefont {A.}~\bibnamefont {Romanenko}},
  \bibinfo {author} {\bibfnamefont {O.}~\bibnamefont {Melnychuk}}, \bibinfo
  {author} {\bibfnamefont {D.}~\bibnamefont {Sergatskov}}, \bibinfo {author}
  {\bibfnamefont {S.}~\bibnamefont {Posen}},\ and\ \bibinfo {author}
  {\bibfnamefont {J.}~\bibnamefont {Zasadzinski}},\ }\bibfield  {title}
  {\enquote {\bibinfo {title} {Effect of interstitial impurities on the field
  dependent microwave surface resistance of niobium},}\ }\href@noop {}
  {\bibfield  {journal} {\bibinfo  {journal} {Applied Physics Letters}\
  }\textbf {\bibinfo {volume} {109}},\ \bibinfo {pages} {062601} (\bibinfo
  {year} {2016})}\BibitemShut {NoStop}%
\bibitem [{Aqu()}]{Aquadag}%
  \BibitemOpen
  \href@noop {} {\enquote {\bibinfo {title} {{
  Aquadag\textsuperscript{\textregistered} E, Aqueous Deflocculated Acheson
  Graphite, Ladd Research, Williston, VT, USA}},}\ }\BibitemShut {NoStop}%
\end{thebibliography}%

\end{document}